\documentclass[twoside]{article}
\usepackage[papersize={5.5in,8.27654in},total={4.25in,7.25in},%
  includehead,nofoot,centering]{geometry}
\usepackage{amsmath,txfonts}
\usepackage[T1]{fontenc}
\usepackage[colorlinks=true,allcolors=blue,bookmarks=false,
  pdfauthor={Gavin R. Putland},
  pdftitle={The unreasonable effectiveness of the cathetus rule in ancient and modern optics},
  pdfkeywords={geometrical optics, Gaussian optics, history of optics, stigmatism, astigmatism, sagittal focus},
  pdfsubject={Optics, Geometrical Optics, History of Optics}]{hyperref}
\usepackage{graphicx}
\usepackage[small]{caption} 

\usepackage[defaultlines=2,all]{nowidow}
\usepackage[final,kerning,spacing]{microtype}
  \SetExtraKerning{encoding={*}}{1={-35,-45}, / ={50,50},
    \textemdash={50,50}, \textendash={30,30}, \textquotedblleft={-30,}}
  \SetExtraSpacing{encoding={*}}{A={-80,,}, f={80,,}, r={-80,,}, v={-80,,},
    w ={-70,,}, y={-60,,}, ' = {-200,,}, \textquotedblright={-80,,}}

\title{\vspace{-2em}The unreasonable effectiveness of the cathetus rule in ancient and modern optics}
\author{Gavin R.{\small~}Putland\,\thanks{\,\small Royal Melbourne Institute of Technology, Australia.~ Gmail name: grputland.~ This paper has been \href{https://en.wikiversity.org/wiki/WikiJournal_Preprints/The_unreasonable_effectiveness_of_the_cathetus_rule_in_ancient_and_modern_optics}{submitted to the \textit{WikiJournal of Science}} under a \href{https://creativecommons.org/licenses/by/4.0/legalcode}{Creative Commons Attribution 4.0 International} license (submission not acknowledged as of date above).}}
\date{\normalsize\vspace{-1ex}27 June 2026} 

\makeatletter
\def\ps@plain{
  \let\@oddfoot=\@empty
  \def\@oddhead{\hfill\normalsize\sf\thepage}
  }
\def\ps@headings{
  \def\sectionmark##1{\markright{\S\thesection.~ ##1}}
  \let\@evenfoot=\@empty  \let\@oddfoot=\@empty
  \def\@evenhead{\footnotesize\sf\underline{\makebox[\textwidth]{\normalsize\sf\thepage\hfill\footnotesize\sf\phantom{g}Putland,\itshape\, The unreasonable effectiveness of the cathetus rule\ldots}}}
  \def\@oddhead{\footnotesize\sf\underline{\makebox[\textwidth]{\rightmark\phantom{g}\hfill\normalsize\sf\thepage}}}
  }
\makeatother

\begin{document}

\sloppy

\maketitle

\vspace{-2em}

{\small
\tableofcontents
}

\clearpage

\pagestyle{headings}

\subsection*{Abstract}
\addcontentsline{toc}{section}{Abstract\vspace{-.8ex}}

The "cathetus rule" in optics alleges that the image of an object-point, formed by reflection or refraction at a surface, lies on the perpendicular ("cathetus") from the object-point to or through the surface. The first known statement of the rule, attributed to Euclid, was for a plane or spherical mirror. The rule was extended to refraction by Ptolemy and to cylindrical and conical mirrors by Ibn al-Haytham, and was upheld by Witelo. But the first valid proofs involving lines of sight other than the cathetus itself were published by Benedetti as late as 1585, for binocular vision, for two special cases: (i) a plane mirror, and (ii) a concave or convex spherical mirror with the two points of reflection (one for each eye) equidistant from the cathetus. Benedetti also gave the first explicit counterexamples to the rule -- for a concave or convex spherical mirror with the eyes in the same plane of reflection. Kepler, in 1604, used more general lines of sight than Benedetti, improved on Benedetti's counterexample for the convex spherical mirror, gave the first counterexample for refraction, salvaged the rule for reflection or refraction in a plane or spherical surface subject to appropriate symmetry in the placement of the eyes, offered the first rebuttals of the received rational arguments for the rule, and did all this in a systematic treatise on "the optical part of astronomy", which so eclipsed Benedetti's book that Kepler was universally credited with the first disproof-and-salvage of the cathetus rule until 2018, when Benedetti's priority was exposed by Goulding.

Kepler notwithstanding, the rule was reaffirmed by Tacquet for plane and spherical mirrors, except for the case in which the rays converge toward a point behind the eye; this became known as the "Barrovian case" because it troubled Barrow, in spite of his modern concept of an image. Barrow demolished the cathetus rule for the tangential image except in the paraxial limit, and Newton salvaged it for the sagittal image. The rule then seems to fade from history.

But the rule is equivalent to the assumption that the image is stigmatic and the cathetus well defined. This narrow assumption is approximately true in the first-order (paraxial, "Gaussian") analysis of lenses and mirrors; and unacknowledged applications of the ancient rule can indeed be discerned in modern expositions of that subject. Moreover, the validity of the rule for the sagittal image fills a critical gap in meridional ray-tracing through spherical surfaces: by tracing the chief ray from an off-axis object-point, then applying the cathetus rule to the successive surfaces, one can locate successive sagittal image-points on the chief ray (produced rectilinearly through surfaces as necessary), and hence assess astigmatism to leading order, without tracing any rays outside the meridional plane.

\clearpage

\section[Introduction: Undeniable implausibility\vspace{-1.2ex}]{Introduction: Undeniable implausibility}

\begin{figure}[h]\centering
\includegraphics[width=3in]{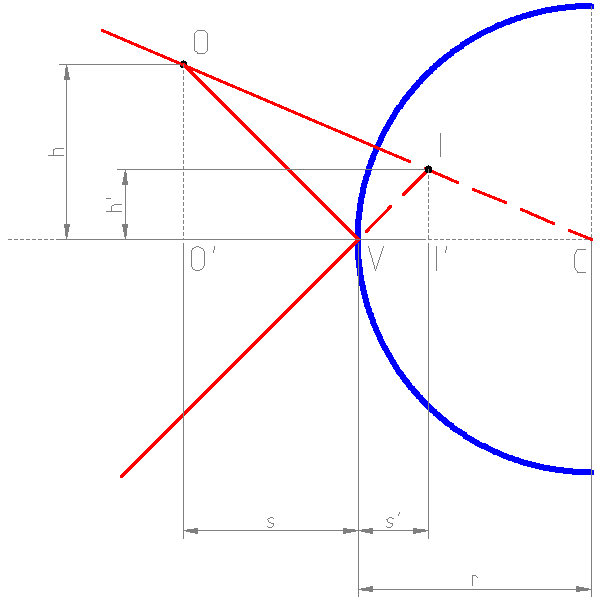}
\caption{This modern diagram, for locating the image~$I$ of an object-point~$O$ in a convex spherical mirror whose center of curvature is~$C$, happens to agree with the ancient cathetus rule. In~this case the cathetus is $\mathit{OC}${\tiny\,} and the point of reflection is~$V$.\, According to the rule, the image is at the intersection of the line of sight (through the point of reflection) and the cathetus. (\href{https://commons.wikimedia.org/wiki/File:Convex_mirror.png}{Diagram by `Forna'} at \textit{Wikimedia Commons}; public domain.)}
\label{f-Forna}
\end{figure}

The \textit{cathetus rule}, as it came to be called, is the ancient optical principle according to which the image of an object-point formed by a reflective or refractive surface lies at the intersection of the line of sight and the \textit{cathetus}, the latter being the perpendicular let fall from the object-point to the surface. The line of sight and/or the cathetus may be produced rectilinearly through the surface. In~the earliest statements of the rule, but not all statements, the surface is assumed to be plane or spherical. If the premise that the image-point lies on the line of sight is taken as tautological, the rule reduces to the proposition that the image-point lies on the cathetus, but still carries the implication that the line of sight intersects the cathetus.

The rule is easily distilled to an absurdity, especially if we drop the assumption that the surface is plane or spherical. \hypertarget{active}{Suppose that the image is seen in a part of the surface (which we shall call the \textit{active} part) far removed from the cathetus.} If we now deform the surface in a small neighborhood of the cathetus so that the cathetus moves, does the image also move although the object and the observer and the active part of the surface do not? Or if, while preserving the active part, we damage another part of the surface so that there is no longer any cathetus, does the image disappear? For that matter, does the image disappear\textemdash even in a plane or spherical surface\textemdash if we merely cover the point on the surface where the cathetus falls?

\section[History\vspace{.3ex}]{History}

\subsection{Euclid}

In the oldest surviving source of the cathetus rule, namely the \textit{\href{https://en.wikipedia.org/wiki/Catoptrics}{Catoptrics}} traditionally attributed to \href{https://en.wikipedia.org/wiki/Euclid}{Euclid}, the last-mentioned absurdity seems to be not only tolerated as an implication, but relied upon as a premise, and even stated among the postulates at the outset: the 4th and 5th postulates, as paraphrased by \href{https://en.wikipedia.org/wiki/A. Mark Smith}{A.\,Mark Smith}, state that in plane, convex spherical, and concave spherical mirrors, ``if a perpendicular (the so-called cathetus) is dropped from an object to the mirror's surface, and if the point at which it meets that surface is covered, the object will no longer be seen.''{\kern.5bp}\footnote{\,\small\hyperlink{smith-2017}{Smith, 2017}, p.\,56. For the original Greek and the Latin translation by Jean Pena, see \hyperlink{euclid-pena-1557}{Euclid/Pena, 1557}, p.\,35 in the Greek version, \&~p.\,45 in the Latin version. Smith evidently follows a different edition in numbering the offending postulates as 4 and 5; though I~have small Greek and less Latin, I~notice that Pena's edition divides the corresponding postulates into nos.\,4,\,5,\,and~6, referring respectively to plane, convex spherical, and concave spherical mirrors.}{\tiny\,} Euclid cites these postulates, together with the premise that the image lies on the line of sight (Postulate~2), to prove the cathetus rule for plane mirrors (Proposition~16), convex spherical mirrors (Proposition~17), and concave spherical mirrors (Proposition~18).\footnote{\,\small See \hyperlink{euclid-pena-1557}{Euclid/Pena, 1557}, p.\,42 in the Greek \&~pp.\,55--6 in the Latin.}

\hypertarget{takahashi-defense}{\href{https://en.wikipedia.org/wiki/Ken'ichi Takahashi}{Ken'ichi Takahashi} has suggested, in Euclid's defense, that the 4th and 5th postulates refer correctly to the case in which the observer looks along the cathetus, so that the line of sight is blocked by the object},\footnote{\,\small\hyperlink{takahashi-92}{Takahashi, 1992}, pp.\,20--26, cited by \hyperlink{smith-2017}{Smith, 2017}, pp.\,59--61, and by \hyperlink{goulding-18}{Goulding, 2018}, pp.\,500--501.} or, I~should add, by the observer's head, if it is between the object and the mirror. Under that interpretation, the cathetus rule seems to be based on the reasonable premise that the image-point lies at the intersection of \textit{two} lines of sight. But that does not explain why the cathetus (if it exists) must be one of them, or why all choices of the other should intersect the cathetus at the same point (if at all), or how we can speak of ``the'' image if they do not. Neither does any ``two lines of sight'' argument appear in subsequent ancient and medieval efforts to defend the rule (as we shall see). Nevertheless the rule is upheld as often as it is mentioned, for both reflection and refraction, by all optical writers until \href{https://en.wikipedia.org/wiki/Giambattista Benedetti}{Benedetti} (1585),\footnote{\,\small\hyperlink{goulding-18}{Goulding, 2018}.} and by all better-known ones until Kepler.\footnote{\,\small\hyperlink{darrigol-12}{Darrigol, 2012}, pp.\,26--7.}

\href{https://en.wikipedia.org/wiki/Johannes Kepler}{Johannes Kepler}, in the third chapter of his \textit{Paralipomena} (1604), initially interprets Euclid's premise in the more literal, absurd manner, and duly dismisses it. Supposing that \textit{C} is the foot of the cathetus from the object-point~\textit{A},\, Kepler says of Euclid:
{\small\begin{quote}
That the place of the image of the object \textit{A} is on \textit{AC}{\tiny\,} he proves thus: ``For,'' he says, ``when the position \textit{C} of the mirror is taken, upon which the perpendicular falls, the visible object \textit{A} is no longer seen.'' If by ``taken'' you understand ``occupied'' (that~is, that the position \textit{C} is covered), the axiom is false\ldots\footnote{\,\small\hyperlink{kepler-donahue-00}{Kepler/Donahue, 2000}, p.\,73. In~the quote, which appears in italics in the original (\hyperlink{kepler-1604}{Kepler, 1604}, p.\,56), Kepler may be translating from Greek, or paraphrasing, rather than quoting from Latin; \textit{cf}.~\hyperlink{euclid-pena-1557}{Euclid/Pena, 1557}, p.\,42 in the Greek \&~p.\,55 in the Latin.}
\end{quote}}\noindent
Kepler offers Euclid a lifeline but cannot save him:
{\small\begin{quote}
Let us now grant that Euclid's axiom is to be understood differently, so as to state that if the observer were situated at \textit{A}\, and \textit{C} were covered, then \textit{A} would not be seen. Then the~axiom is perfectly true, but the conclusion does not follow from it, except for perpendicular viewing. The argument does not carry over from a perpendicular to an oblique observer.\footnote{\,\small\hyperlink{kepler-donahue-00}{Kepler/Donahue, 2000}, p.\,74.}
\end{quote}}\noindent
Kepler's lifeline is not as general as Takahashi's; but even if it were, the argument would still ``not carry over'' to an oblique viewer, in as much as it would not explain why the line of sight should intersect the cathetus.

In Euclid's Postulate~4 and Proposition~16, the Greek \textit{k\'{a}thetos} is rendered in Latin as \textit{perpendicularis} by at least three translators,\footnote{\,\small\hyperlink{euclid-pena-1557}{Euclid/Pena, 1557}, pp.\,45 \&~55 in the Latin; \hyperlink{euclid-dasypodius-1557}{Euclid/Dasypodius, 1557} (unnumbered pages); \hyperlink{euclid-heiberg-1895}{Euclid/Heiberg, 1895}, pp.\,286--7,\,312--13.} whereas Postulate~5 and Propositions 17 and~18 refer to the cathetus not by any name, but as the line drawn to the center of the sphere.

\subsection{Ptolemy}

When interpreting the authorities on geometrical optics before 1000{\footnotesize\,CE}, we must remember that they believed in visual rays emitted by the eye, so that the ``incident'' ray is from the eye, not from the object-point; the ``cathetus of incidence'' (if~it is mentioned) is therefore the perpendicular from the \textit{eye} to the surface, while the usual ``cathetus'' (the perpendicular from the \textit{object-point} to the surface) may be called the cathetus of reflection or refraction. So it is with \href{https://en.wikipedia.org/wiki/Ptolemy}{Ptolemy}'s \textit{Optics}, written some years after his \textit{\href{https://en.wikipedia.org/wiki/Almagest}{Almagest}}, but known to us only through a 12th-century Latin translation of a now lost, incomplete Arabic translation.\footnote{\,\small\hyperlink{smith-1996}{Smith, 1996}, pp.\,1--8; \hyperlink{lindberg-81}{Lindberg, 1981}, p.\,211.} (Even the Latin version was not available to Kepler,\footnote{\,\small\hyperlink{kepler-donahue-00}{Kepler/Donahue, 2000}, p.\,84, n.\,34; \hyperlink{lohne-59}{Lohne, 1959}, pp.\,117-18.} and not printed until 1885.\footnote{\,\small\hyperlink{ptolemy-govi-1885}{Ptolemy/Govi, 1885}.})

Ptolemy affirms the cathetus rule for reflection in a plane or spherical mirror, on the empirical ground that a thin rod standing perpendicularly on the reflecting surface appears aligned with its reflection\footnote{\,\small\hyperlink{smith-2017}{Smith, 2017}, p.\,93.} when ``properly viewed outside the mirror.''{\kern.5bp}\footnote{\,\small\hyperlink{smith-1996}{Smith, 1996}, pp.\,131--2.}{\tiny\,} That premise is certainly true if the rod is viewed with one eye, due to the axial symmetry about the rod (the cathetus), implying a bilateral symmetry (``mirror symmetry'') about the plane of the eye and the rod. But it proves only that the image is in that plane\textemdash not that it is necessarily collinear with the rod. Moving the eye around the rod does not prove anything more, because the said plane moves with the eye, so that the image, if not collinear with the rod, moves with the plane.

Ptolemy then notes that the perpendicular to the surface at the point of reflection is in the plane of the line of sight and the cathetus,\footnote{\,\small\hyperlink{smith-1996}{Smith, 1996}, p.\,132.} which is indeed the case if we retain the symmetry. Thus he makes the cathetus rule the \textit{premise} of an aspect of the law of reflection\textemdash an aspect that seems to have escaped his predecessors\footnote{\,\small\hyperlink{smith-1996}{Smith, 1996}, p.\,36.}\textemdash\,namely that the incident and reflected rays and the normal at the point of reflection are coplanar!{\kern.5bp}\footnote{\,\small\textit{Cf}.~\hyperlink{goulding-18}{Goulding, 2018}, p.\,502.}

\hypertarget{floating-coin}{Later in his treatise, Ptolemy makes the corresponding aspect of the law of \textit{refraction} dependent on the cathetus rule. As evidence for the latter, he cites the already old ``floating coin'' experiment,} in which a coin lying on the bottom of a tub and hidden by the rim is seemingly raised into view by filling the tub with water.\footnote{\,\small\hyperlink{smith-1996}{Smith, 1996}, pp.\,230--31.} He does not explain why the image should be raised precisely \textit{vertically}, as the cathetus rule requires\textemdash and as will \textit{appear} to be confirmed in observations that tacitly exploit the axial symmetry. And although the cited experiment concerns a \textit{plane} refracting surface, Ptolemy goes on to apply the rule to spherical refracting surfaces without further justification.\footnote{\,\small\hyperlink{smith-1996}{Smith, 1996}, pp.\,252--3.}

In addition to these flawed empirical demonstrations of the rule, Ptolemy attempts a rational explanation, saying that the location of the image must be unique, and that ``to any point on a given object there is one and only one cathetus, whereas any other line, being oblique with respect to this cathetus, is subject to numerous variations.''{\kern.5bp}\footnote{\,\small Translated by Smith (\hyperlink{smith-1996}{1996}, p.\,138); cited by Goulding (\hyperlink{goulding-18}{2018}, p.\,503).}{\tiny\,} There are at least two weaknesses in this argument. First, some qualification must be imposed on the image-point in order to ensure uniqueness; Ptolemy himself shows that for given positions of the object-point and the eye, a concave mirror can give multiple points of reflection, which, according to the cathetus rule, will give multiple image-points on a common cathetus.\footnote{\,\small\hyperlink{smith-2017}{Smith, 2017}, p.\,102; \hyperlink{goulding-18}{Goulding, 2018}, p.\,505\textit{n}.} Second, and more seriously, even if the image-point and the cathetus are both unique, that does not prove any other connection between the two!

In the Latin text of Ptolemy's \textit{Optics}, which is already a translation of a translation, the cathetus is again called the \textit{perpendicularis}.\footnote{\,\small\hyperlink{smith-1996}{Smith, 1996}, pp.\,287--8, 296. The word \textit{cathetus} and the expressions \textit{cathetus of incidence} and \textit{cathetus of reflection} appear in Smith's English translation, and these terms together with \textit{cathetus of refraction} appear in his annotations.}

\medskip

In antiquity, the cathetus rule was found effective in~spite of its lack of foundation, and not only for establishing the coplanarity laws of reflection and refraction. Its effectiveness for \textit{reflection} is accidentally emphasized by one medieval author who seems unfamiliar with the rule: the Syrian Christian polymath \href{https://en.wikipedia.org/wiki/Qusta ibn Luqa}{Qus\d{t}\={a} ibn L\={u}q\={a}} (820?--912?{\footnotesize\,CE}). In~only one case\textemdash that of a plane mirror\textemdash does he specify the location of a reflected image. To explain why~(e.g.) the image in a convex mirror is diminished,\footnote{\,\small\hyperlink{smith-2017}{Smith, 2017}, pp.\,170--71.} Ibn\,L\={u}q\={a} compares the apparent extent of the image \textit{on the reflecting surface} with that given by a plane mirror\textemdash whereas Euclid\footnote{\,\small\hyperlink{smith-2017}{Smith, 2017}, p.\,61.} and Ptolemy,\footnote{\,\small\hyperlink{smith-1996}{Smith, 1996}, pp.\,165--9.} aided by the cathetus rule, have correctly deduced not only that the image is diminished, but also that it is closer to the reflective surface than the object is, and that convex mirrors make the world look convex.

By the end of the 10th century, however, Ptolemy's \textit{Optics} has been translated into Arabic,\footnote{\,\small\hyperlink{smith-1996}{Smith, 1996}, p.\,6.} ready to be studied\textemdash and surpassed\textemdash by ``the most significant figure in the history of optics between antiquity and the seventeenth century.''{\kern.5bp}\footnote{\,\small\hyperlink{lindberg-81}{Lindberg, 1981}, p.\,58.}

\subsection{Alhacen}

Ab\={u} `Al\={\i} al-\d{H}asan (``Alhacen'') ibn al-\d{H}asan \href{https://en.wikipedia.org/wiki/Ibn al-Haytham}{ibn al-Haytham}\footnote{\,\small{}The original Latinization of his name was \textit{Alhacen}, not the more familiar \textit{Alhazen} (\hyperlink{lindberg-81}{Lindberg, 1981}, pp.\,209--10; \hyperlink{smith-2017}{Smith, 2017}, p.\,1).} wrote his \textit{Book of Optics} circa~1030{\footnotesize\,CE}.\footnote{\,\small\hyperlink{smith-2017}{Smith, 2017}, p.\,182.} For Alhacen, the eye is not an emitter of visual rays, but a receiver of light rays.\footnote{\,\small\hyperlink{darrigol-12}{Darrigol, 2012}, pp.\,17--18; \hyperlink{smith-2017}{Smith, 2017}, pp.\,184--6.}$^,$\footnote{\,\small{}Although Alhacen's theory of vision was not the first \textit{intromission} theory, it was apparently the first such theory to incorporate the premise (first stated explicitly by \href{https://en.wikipedia.org/wiki/al-Kindi}{al-Kind\={\i}} in the 9th century) that each visible spot on a luminous or illuminated body sends out \textit{light}, and consequently the first such theory that could be reconciled with a geometrical science of optics (\hyperlink{lindberg-81}{Lindberg, 1981}, pp.\,30--31,\,58--60).} Hence, in reflection or refraction, the ``incident'' ray is not from the eye, but from the object-point, and the ``perpendicular of incidence'' is dropped from the object-point, while the ``line of sight'' now coincides with the ``line of reflection'' or the ``line of refraction''\!. This reversal of direction does not affect the geometry and therefore does not of itself furnish any new arguments for the cathetus rule, although Alhacen offers many\textemdash some empirical and some rational, for both reflection\footnote{\,\small\hyperlink{smith-2006}{Smith, 2006}, pp.\,385--97.} and refraction\footnote{\,\small\hyperlink{smith-2010}{Smith, 2010}, pp.\,275--82.}\textemdash none of which is an exemplar of the rigor for which he is otherwise renowned.

In the \textit{empirical} category, for a plane mirror,\footnote{\,\small\hyperlink{smith-2006}{Smith, 2006}, pp.\,385--7.} Alhacen recommends putting marks on Ptolemy's rod (but does not name Ptolemy here). Then he tries a cone instead of a rod, and invites us to imagine such a cone extended to the mirror from every point on the object. He notes that the same observations hold for convex spherical mirrors.\footnote{\,\small\hyperlink{smith-2006}{Smith, 2006}, pp.\,387--8.} Conceding that they do \textit{not} generally hold for a convex \textit{cylindrical} mirror, because ``what is straight does not appear straight''\!, Alhacen claims that the cathetus rule is still verified for a \textit{single point} on the object seen in such a mirror.\footnote{\,\small\hyperlink{smith-2006}{Smith, 2006}, p.\,388.} It~seems to escape his notice that if the image of a point on a thin rod standing perpendicularly on the mirror does not align with the rod, then the line of reflection, when produced through the mirror, does not intersect the cathetus at~all. Obviously, by symmetry, the image will appear to align with the rod if the plane of the eye and the rod contains the axis of the cylinder or is perpendicular thereto; and his experiments confirm these cases.\footnote{\,\small\hyperlink{smith-2006}{Smith, 2006}, pp.\,388--9.} In intermediate cases, if the image of the tip of the rod is to fall on the cathetus, the line of sight and therefore the point of reflection must be in the plane of the eye and the cathetus, so that the point of reflection must be on the elliptical section of that cylinder by the plane\textemdash which is precisely what Alhacen claims,\footnote{\,\small\hyperlink{smith-2006}{Smith, 2006}, pp.\,389--91 (pars.\,2.15--18) and note~12 (p.\,489), referring to figure~5.2 on p.\,216 (other~volume).} without checking the requirement that the normal to the cylinder at this point is in the same plane (that of the incident and reflected rays), as he stipulates in his statement of the law of reflection.\footnote{\,\small\hyperlink{smith-2006}{Smith, 2006}, p.\,300.}

After briefly claiming that the same procedure can be applied to convex conical mirrors, with the same results~(!), Alhacen turns to concave spherical mirrors.\footnote{\,\small\hyperlink{smith-2006}{Smith, 2006}, pp.\,391--4.} Fashion a right circular cone whose slant height is equal to the radius of curvature of the mirror, mark a ``line of longitude'' (generating line) on the cone, and mount the cone on the mirror, so that the apex of the cone is at the center of curvature of the mirror; then, he says, the cone and the line of longitude will appear to extend into the mirror. Next, having placed the apex at the center of curvature, mount a thin rod on the mirror so that its tip is between the apex and the mirror while the image of the tip is in front of the mirror; then the image will be nearer to the eye (note the singular) than the apex is, and you will be able to bring the tip, the apex, and the image into a single line of sight. Finally he claims that the cathetus rule holds for concave cylindrical and conical mirrors, by the same flawed reasoning as for their convex counterparts.

In the account of the concave spherical mirror, modern readers will recognize the apparent continuation of the cone into the mirror as the virtual image of an object inside focus, and will recognize the image of the tip of the rod as the real image of an object-point between the focus and the center of curvature. Otherwise the above observations of Alhacen, in so far as they are correct, are trivial consequences of the axial symmetry of the surface about the cathetus or catheti; and in only one case\textemdash that in which we look along the cathetus, through the image of the rod-tip to the tip itself\textemdash does he establish that the image is on the cathetus and not merely in the plane of the cathetus \& the eye.

For refraction, Alhacen rightly cites the \hyperlink{floating-coin}{floating-coin experiment} as proof that the image is displaced from the object.\footnote{\,\small\hyperlink{smith-2010}{Smith, 2010}, pp.\,274--5.} He then asserts the cathetus rule, and claims to prove it by a variant experiment in which a vertical diameter and a sloping diameter are marked on a vertical disk, which is immersed in water up to a point above the intersection (center of the disk), with the marked surface facing the eye (note the singular), which is best placed just above the water level. The vertical line then appears to continue vertically into the water, so that the point of intersection (the object-point) appears to lie on the continuation (the cathetus), while the sloping line appears to be kinked at the surface. Alhacen further recommends rotating the disk so as to interchange the roles of the two marked diameters.\footnote{\,\small\hyperlink{smith-2010}{Smith, 2010}, pp.\,275--7.} But, whichever diameter is the cathetus, he again fails to explain why the image is on the cathetus and not merely in the plane of the cathetus and the eye.

\begin{figure}[t]\centering
\includegraphics[width=2.56in]{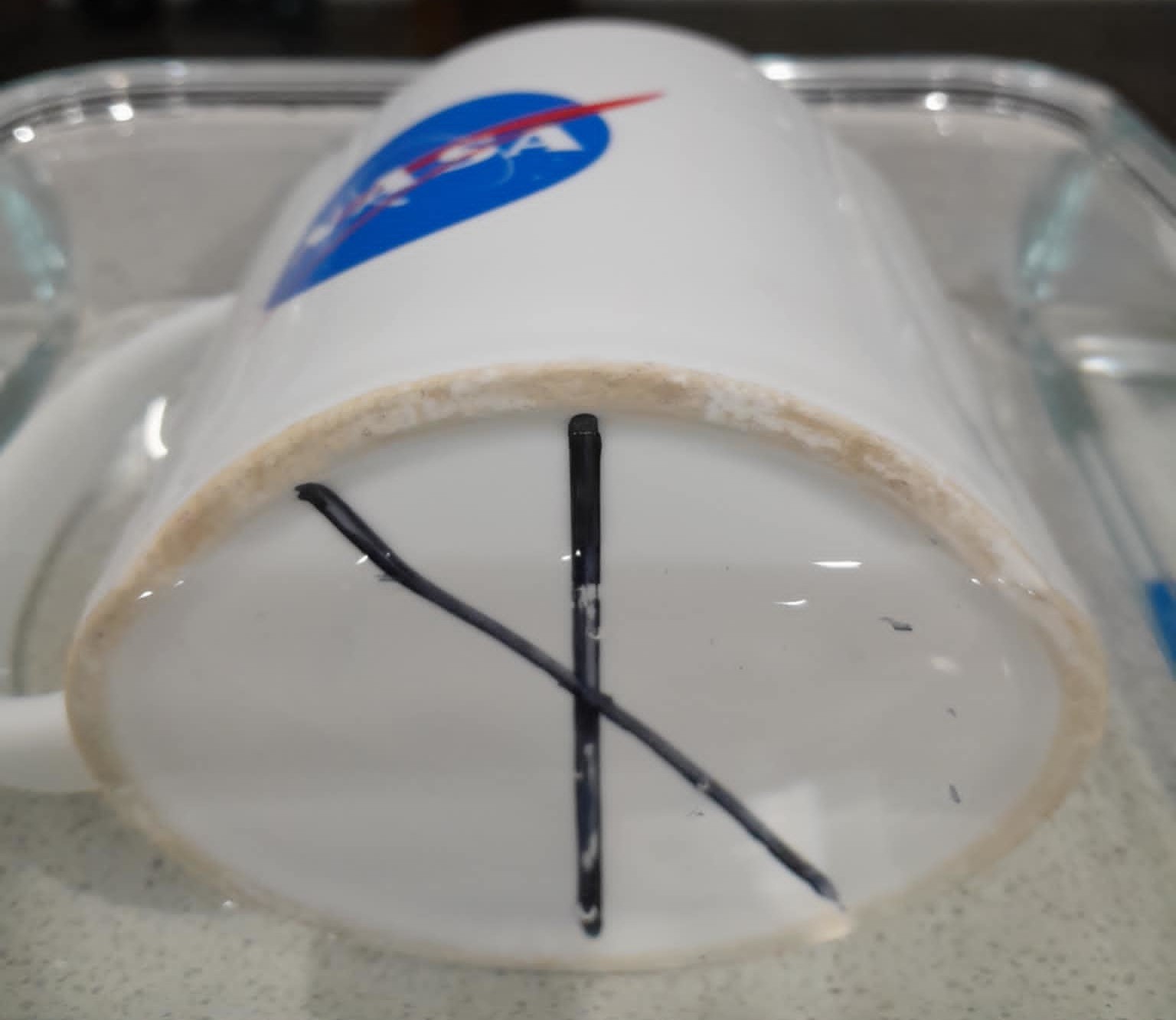}
\caption{Kitchen-bench reconstruction of Alhacen's first experiment attempting to prove the cathetus rule for refraction. A~vertical diameter and a sloping diameter are drawn on the base of a coffee mug. The vertical diameter appears to continue vertically into the water, showing that the \textit{image} of the point of intersection is in the plane of the viewing position and the cathetus (vertical diameter). Alhacen would claim that the image-point is not~only in this plane, but \textit{on} the cathetus. (Photo by the author; public domain.)}
\end{figure}

This defect is not repaired by the next experiment, using the same disk but no water, which is intended to interchange the places of the rare and dense media.\footnote{\,\small\hyperlink{smith-2010}{Smith, 2010}, pp.\,277--80.} A~rectangular glass block, with its top and bottom faces horizontal, is affixed to the disk near the top, covering a portion of each marked diameter. The observer's eyes are positioned so that one eye is close to the top face of the block and sees both diameters through the block, while the other eye sees the intersection without refraction (bypassing the block). Then the former eye perceives the entire vertical diameter (the cathetus) as vertical and aligned with the portion seen by the other eye without refraction, although the two eyes see the intersection at different points on the cathetus. Thus the image of the intersection, as seen by the former eye, appears to be on the cathetus. But again this appearance follows from the weaker condition that each eye perceives the vertical diameter (or the relevant part thereof) to be in the \textit{plane} of that eye and the vertical diameter: as the two planes intersect on the vertical diameter, that diameter must appear in its true alignment, even if the eyes disagree on the positions of its constituent points (only one of which\textemdash the intersection\textemdash looks different from the others).

In the \textit{rational} category, for reflection, Alhacen sets out to explain ``why visible objects are perceived through reflection where the image is located and why the image lies on the normal from the visible object to the surface of the mirror.''{\kern.5bp}\footnote{\,\small\hyperlink{smith-2006}{Smith, 2006}, p.\,394.}{\tiny\,} On the latter question, he first says that we judge the distance of an image by comparing its angular size with its absolute size.\footnote{\,\small\hyperlink{smith-2006}{Smith, 2006}, p.\,395.} For the purpose of establishing the cathetus rule, by which we propose to locate points on images and thence determine absolute sizes of images, this is a circular argument.

For plane mirrors, says Alhacen, ``since the image does not appear on the surface of the mirror but behind it, it is more appropriate and reasonable for it to appear upon rather than outside the normal.''{\kern.5bp}\footnote{\,\small\hyperlink{smith-2006}{Smith, 2006}, p.\,395; \textit{cf}.~\hyperlink{goulding-18}{Goulding, 2018}, p.\,504.}{\tiny\,} Taking that as a \textit{premise}, he correctly locates the image. He adds that if the image were beyond or in front of the cathetus, then, since the image lies on the line of reflection, it would be further from or nearer to the eye and would therefore subtend a smaller or larger angle.\footnote{\,\small\hyperlink{smith-2006}{Smith, 2006}, p.\,396.} But in fact, according to the law of reflection, it would subtend the \textit{same} angle because the line of reflection from each point on the object would be unchanged. Kepler raises another objection: Alhacen ``says that when an image is perceived on the perpendicular, it has the proper magnitude belonging to the thing itself.'' But this magnitude, as Kepler notes, cannot be a necessary condition for the correct location of the image, because it does not hold for curved mirrors.\footnote{\,\small\hyperlink{kepler-donahue-00}{Kepler/Donahue, 2000}, p.\,75.}

For a convex mirror, Alhacen argues verbosely but validly that the image of the center of the eye (note the singular) must be on the cathetus due to symmetry. But then he extends the argument to the image of any other point on the eye, although the symmetry is broken in that the image is no longer \textit{seen} along the cathetus; and he briefly claims that the same logic applies to a concave spherical mirror and to a concave or convex conical mirror,\footnote{\,\small\hyperlink{smith-2006}{Smith, 2006}, pp.\,396--7.} although in the conical case, even the surface is not axially symmetrical about the cathetus.

Just before the claim on concave and conical mirrors, Alhacen says in support of the cathetus rule:
{\small\begin{quote}
\hypertarget{alhacen-obj-img}{The state of natural things is in accordance with the situation of their principles}, and the principles of natural things are hidden.\footnote{\,\small Quoted in translation by Goulding (\hyperlink{goulding-18}{2018}, p.\,505); \textit{cf}.~\hyperlink{smith-2006}{Smith, 2006}, p.\,397.}
\end{quote}}\noindent
``By these words he says two things,'' says Kepler. ``First, he repeats the very thing that was proposed to prove (for they say nothing different), and second, he says by way of appending the cause, that it is hidden. But this is not demonstrating.''{\kern.5bp}\footnote{\,\small\hyperlink{kepler-donahue-00}{Kepler/Donahue, 2000}, pp.\,74--5.}{\tiny\,} And just \textit{after} the claim on concave and conical mirrors, Alhacen continues:
{\small\begin{quote}
And the place of the image will universally be on the perpendicular in any mirror, because there is no place outside the perpendicular in which the form maintains a likeness and identity of position.\footnote{\,\small\hyperlink{goulding-18}{Goulding, 2018}, p.\,505; \textit{cf}.~\hyperlink{smith-2006}{Smith, 2006}, p.\,397.}
\end{quote}}\noindent
Thus he seems to argue from the location of the thing seen to the location of the image; this mode of reasoning will reappear later.

Broadening the attack, Kepler adds: ``\textit{But this fact further strongly confutes the Optical writers}, that they do not give the same cause of this matter in reflection as in refraction.''{\kern.5bp}\footnote{\,\small Italics in the Latin (\hyperlink{kepler-1604}{Kepler, 1604}, p.\,58), not quite matching \hyperlink{kepler-donahue-00}{Kepler/Donahue, 2000}, p.\,75.}{\tiny\,} Indeed, in support of the cathetus rule for refraction, Alhacen apparently reasons that the motion of the light ray in the medium containing the object-point can be resolved into a component in the direction of the cathetus, and a component perpendicular thereto.\footnote{\,\small\hyperlink{smith-2010}{Smith, 2010}, pp.\,280--82.} An obvious weakness in that argument, if we credit it with any relevance at all (which Kepler does not), is that we can choose the former direction differently and still perform the resolution. Kepler also argues, somewhat cryptically, that refraction further weakens Alhacen's connection between image size and correct image location, in~that the size-distance relation for refraction is different from that for reflection.\footnote{\,\small\hyperlink{kepler-donahue-00}{Kepler/Donahue, 2000}, p.\,76.}

On three pillars\textemdash the cathetus rule, the correct law of reflection, and an incomplete law of refraction\textemdash Alhacen builds a comprehensive and largely correct theory of image location, magnification, and distortion in seven types of mirrors,\footnote{\,\small\hyperlink{smith-2017}{Smith, 2017}, pp.\,204--5.} image location in plane and spherical refracting surfaces,\footnote{\,\small\hyperlink{smith-2010}{Smith, 2010}, chap.\,5, pars.\,25--90 (p.\,282ff).} and magnification by spherical refracting surfaces.\footnote{\,\small\hyperlink{smith-2010}{Smith, 2010}, chap.\,7.} Taking the first pillar to imply that an object-point is perceived to lie on the cathetus from that point to the outer refracting surface of the \textit{eye}, he even offers an explanation why the eye perceives the direction of the object-point although light from that point strikes all points of the eye.\footnote{\,\small\hyperlink{smith-2010}{Smith, 2010}, pp.\,303--4, pars.\,6.22--3. \textit{Cf}.~\hyperlink{lindberg-81}{Lindberg, 1981}, pp.\,76--78. Remarkably, this is the only context in which Lindberg (1981) mentions the cathetus rule (which he states but does not name). More remarkably, he says that in general the rule ``makes perfect sense, for it requires simply that the eye be unaware of the break in the ray\ldots\,and therefore that it project the image backward along the incident ray'' (p.\,76). Apart from misidentifying the ray (an obvious and temporary slip), this explanation fails to explain \textit{how far} the image should be projected back.}$^,$\footnote{\,\small{}But Alhacen does not question the ancient, erroneous doctrine that the glacial humor (lens) is the sensitive part of the eye (\hyperlink{smith-2001}{Smith, 2001}, p.\,417, par.\,2.1). Nor does he deduce (as he would in any other context) that the image-point lies behind the center of the eye (as it does), because that would give an inverted image (as it does), which apparently would imply that we see upside-down! Instead, he concludes that there must be a diverging refraction at the back surface of the glacial humor, so that the cathetal rays from the various object-points do not cross each other (\hyperlink{smith-2001}{Smith, 2001}, pp.\,419--20). \textit{Cf}.~\hyperlink{lindberg-81}{Lindberg, 1981}, pp.\,76--78,\,80--81.} Yet neither he nor anyone before him has offered a firm foundation for that first pillar.

\hypertarget{ingredients}{For the case of reflection in a plane mirror, however, the \textit{ingredients} of a valid proof of the cathetus rule have been unwittingly served up by Ptolemy and Alhacen.} From the law of reflection \textit{and the cathetus rule}, Ptolemy proves that the image-point is as far behind the mirror as the object-point is in front.\footnote{\,\small\hyperlink{smith-1996}{Smith, 1996}, pp.\,155--6 (Theorem~III.5), summarized in \hyperlink{darrigol-12}{Darrigol, 2012}, pp.\,13--14.} If the cathetus rule is not assumed \textit{a~{\tiny\!}priori}, the same geometric argument simply shows that the reflected line of sight to the object-point, when produced from the eye through the mirror, intersects the cathetus as far behind the mirror as the object-point is in front (provided that the line of sight intersects the cathetus at all, as is obvious from the symmetry). By the generality of this line of sight, all such lines of sight intersect the cathetus at the same point, and therefore intersect each other at a common point\textemdash a~\textit{\href{https://en.wikipedia.org/wiki/Stigmatism}{stigmatic}} image\textemdash which is \textit{on the cathetus}. But Ptolemy does not package the argument that way. Nor does Alhacen, who again shows that the line of sight intersects the cathetus as far behind the mirror as the image-point is in front.\footnote{\,\small\hyperlink{smith-2006}{Smith, 2006}, p.\,399 (pars.\,2.47--8 in Prop.\,4).}

\begin{figure}[t]\centering
\includegraphics[width=2.7in]{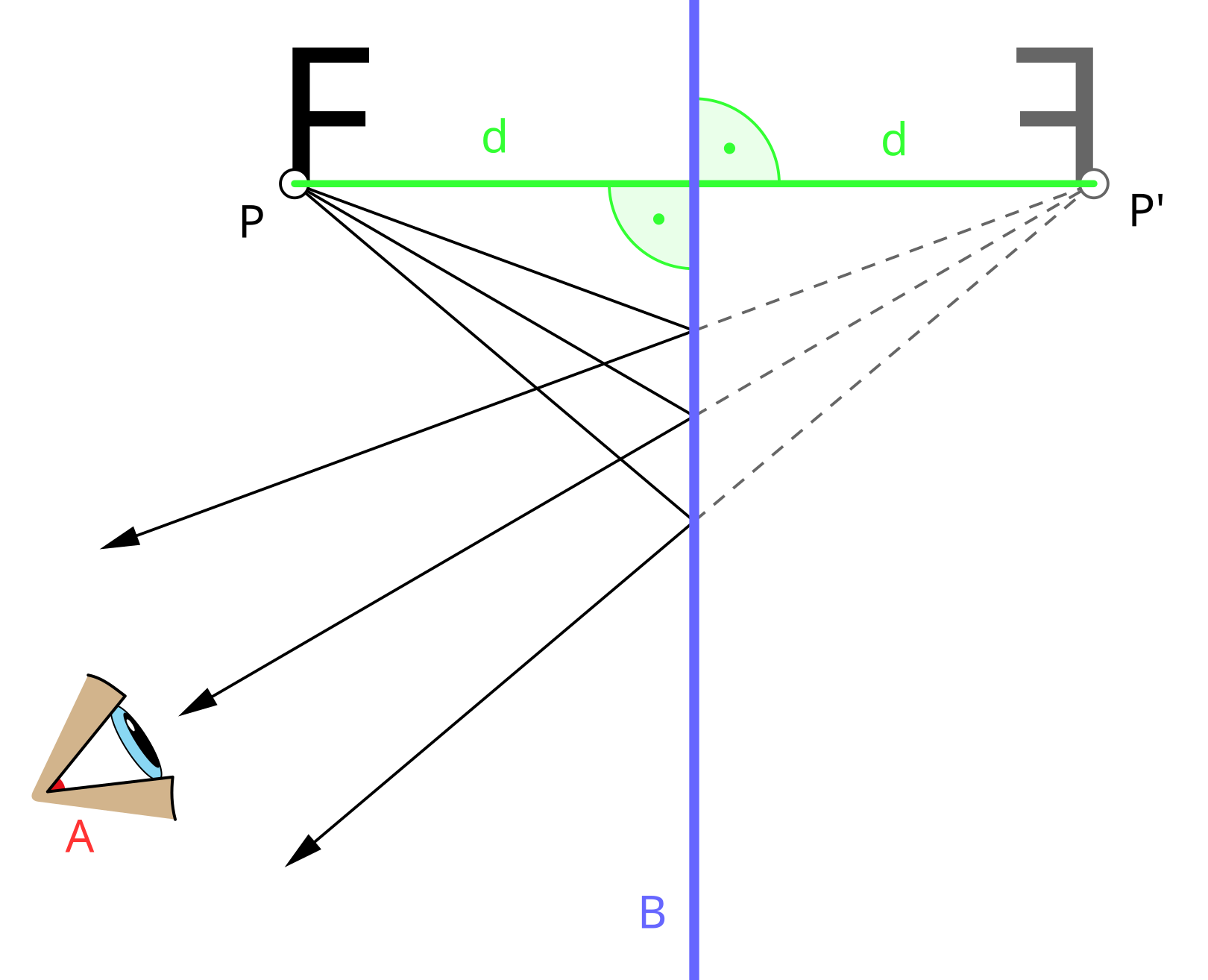}
\caption{Location of the image~$P'$ of an object-point~$P${\tiny\,} in a plane mirror~$B$.\, In~this special case the cathetus rule follows simply and rigorously from the law of reflection (although Ptolemy and Alhacen still cite the cathetus rule independently). (\href{https://commons.wikimedia.org/wiki/File:Plane-mirror_opt.svg}{Original diagram by `MikeRun'} at \textit{Wikimedia Commons}.)}
\end{figure}

\subsection{The three friars}

In the West, as \href{https://en.wikipedia.org/wiki/David C. Lindberg}{David C.~{\tiny\!}Lindberg} explains,
{\small\begin{quote}
the character of the twelfth-century revival of learning was dramatically transformed by a flood of translations from both Greek and Arabic; what was at first chiefly an intensification of interest in ancient Latin sources became a quest for new knowledge, previously unavailable in the West.\,\ldots~In~optics,\,\ldots\,it was not until the middle of the thirteenth century that the full corpus of Greek and Arabic works on the subject was at hand in the major European centers of learning, able to shape (and indeed revolutionize) the thought of Western scholars.\footnote{\,\small\hyperlink{lindberg-81}{Lindberg, 1981}, pp.\,102--3.}
\end{quote}}\noindent
Foremost in the ``corpus'' is Alhacen's \textit{Book of Optics}, translated into Latin circa 1200 as \textit{De\,Aspectibus}. This is the main source, but not the only source, for the three leading Western ``perspectivist''{\kern.5bp}\footnote{\,\small The term was coined by Lindberg (\hyperlink{lindberg-81}{1981}, p.\,251, n.\,1) from late medieval Latin.}{\tiny\,} works, namely
\begin{itemize}\itemsep=-0.4ex
\item\href{https://en.wikipedia.org/wiki/Roger Bacon}{Roger Bacon}'s \textit{Perspectiva}, written circa 1263, and dispatched to the papal court as part~5 of his \textit{Opus Majus} in 1267 or 1268,
\item\href{https://en.wikipedia.org/wiki/Vitello}{Witelo}'s \textit{Perspectiva}, written at the papal court, probably in the first half of the 1270s, and
\item\href{https://en.wikipedia.org/wiki/John Peckham}{John Pecham}'s \textit{Perspectiva Communis}, probably written at the papal court in the late 1270s, just before the author's appointment as Archbishop of Canterbury.\footnote{\,\small\hyperlink{lindberg-71}{Lindberg, 1971}, pp.~{\tiny\!}68--9,\,71 (on~Bacon), 72--3 (on~Witelo), 82--3 (on~Pecham).}
\end{itemize}

Bacon, according to Lindberg, is the first Western optical writer to cite Ptolemy's \textit{Optics}, and only the third to use Alhacen's \textit{De\,Aspectibus}.\footnote{\,\small\hyperlink{lindberg-81}{Lindberg, 1981}, p.\,253, n.\,28.} He also draws on Euclid's \textit{Catoptrics} in a circular attempt to establish the cathetus rule, which he then applies in selected cases.\footnote{\,\small\hyperlink{smith-2017}{Smith, 2017}, pp.\,267--8.} Thus he becomes, as far as I~have noticed in this brief inquiry, the first author to use the Latin term \textit{cathetus} in the optical sense\textemdash mostly in the phrase \textit{cum~catheto} (``with the cathetus'').\footnote{\,\small\hyperlink{bacon-combach-1614}{Bacon/Combach, 1614}, is digitally searchable.}

Witelo is clearly familiar with Bacon's work, presumably through the patronage of the papal confessor (and prolific translator of ancient Greek treatises), \href{https://en.wikipedia.org/wiki/William of Moerbeke}{William of Moerbeke}.\footnote{\,\small\hyperlink{lindberg-71}{Lindberg, 1971}, pp.\,72--5.} But, whereas Bacon summarizes \textit{De\,Aspectibus}, Witelo expands on it, incorporating material from Euclid, \href{https://en.wikipedia.org/wiki/Hero of Alexandria}{Hero of Alexandria}, Ptolemy, al-Kind\={\i}, Alhacen's treatise on parabolic burning mirrors, and \href{https://en.wikipedia.org/wiki/Ibn Mu'adh al-Jayyani}{Ibn\,Mu`\={a}dh}'s essay on twilight, rearranging the content with a mathematical introduction and a consistent theorem-and-proof format\textemdash suitable for a textbook or reference\textemdash and adding a theological prologue for a Roman Catholic readership.\footnote{\,\small\hyperlink{smith-2017}{Smith, 2017}, pp.\,273--5; \hyperlink{unguru-72}{Unguru, 1972}.} And whereas the Latin text of Alhacen's \textit{De\,Aspectibus} does not seem to contain the word \textit{cathetus} or any inflected form thereof (although \textit{perpendicularis} and \textit{perpendiculari} are ubiquitous), Witelo's \textit{Perspectiva} uses that word in some form more than 150 times, including at least 19 occurrences of the phrase \textit{cum~catheto}.\footnote{\,\small\hyperlink{risner-1572}{Risner, 1572}, is digitally searchable.}

Pecham also is clearly familiar with Bacon's work, probably through personal acquaintance, both men having joined the \href{https://en.wikipedia.org/wiki/Franciscans}{Franciscan} order at Oxford in the 1250s and resided at the Franciscan convent in Paris in the 1260s.\footnote{\,\small\hyperlink{lindberg-71}{Lindberg, 1971}, pp.\,75--7.}$^,$\footnote{\,\small{}Witelo is thought to have joined the \href{https://en.wikipedia.org/wiki/Premonstratensians}{Premonstratensians} (Norbertines) in his retirement. Moerbeke was a \href{https://en.wikipedia.org/wiki/Dominican Order}{Dominican}.} Pecham, like Bacon, summarizes Alhacen, but follows him more closely,\footnote{\,\small\hyperlink{smith-2017}{Smith, 2017}, pp.\,273--5.} and again uses the expressions \textit{cathetus} and \textit{cum~catheto}.\footnote{\,\small\hyperlink{pecham-gaurico-1504}{Pecham/Gaurico, 1504}, and \hyperlink{pecham-hartmann-1542}{Pecham/Hartmann, 1542}, are digitally searchable.}$^,$\footnote{\,\small{}Lindberg (\hyperlink{lindberg-71}{1971}, pp.\,66,\,77--83) offers evidence that Pecham was also indebted to Witelo through Moerbeke, but notes that the citations of Witelo in the \textit{Perspectiva Communis} are spurious: they were added by \href{https://en.wikipedia.org/wiki/Georg Hartmann}{Georg Hartmann}, editor of the \hyperlink{pecham-hartmann-1542}{1542 reprint}.}

Bacon's work, although the first of the three to be written, was the last to be printed, in~1614. Pecham's \textit{Perspectiva Communis}, although the last to be written, spawned the largest number of manuscripts, was printed earliest (1482/3) and most often, and was clearly intended for the widest readership;\footnote{\,\small\hyperlink{smith-2017}{Smith, 2017}, p.\,328; Lindberg, \hyperlink{lindberg-81}{1981}, pp.\,120--21.} ``if it were published today,'' says Smith, it ``would probably be retitled \textit{Perspectiva ad asinos} or \textit{Optics for Dummies}.''{\kern.5bp}\footnote{\,\small\hyperlink{smith-2017}{Smith, 2017}, p.\,282.}{\tiny\,} Witelo's \textit{Perspectiva} was printed in 1535 and reissued in 1551. In~1572 it was printed for the third time, and Alhacen's \textit{De\,Aspectibus} for the first time, in~a single weighty volume under the title \textit{Opticae\,Thesaurus}, expertly edited\textemdash reconstructing diagrams and adding explanatory notes, citations of mathematical sources, proposition numbers and headings for Alhacen's work, and cross-references within and between the two works\textemdash by the mathematician \href{https://en.wikipedia.org/wiki/Friedrich Risner}{Friedrich Risner}.\footnote{\,\small\hyperlink{smith-2017}{Smith, 2017}, pp.\,328--9. Smith's translation of Alhacen (\hyperlink{smith-2001}{Smith, 2001}, 2006, 2008, 2010) omits Risner's headings and uses a different section-numbering system.}$^,$\footnote{\,\small{}In the \textit{Opticae\,Thesaurus} (\hyperlink{risner-1572}{Risner, 1572}), the two major treatises are separately paginated. Appended to Alhacen's treatise, at pp.\,283--8, is Ibn~Mu`\={a}dh's essay on twilight\textemdash translated into Latin by \href{https://en.wikipedia.org/wiki/Gerard of Cremona}{Gerard of Cremona} as \textit{De~Crepusculis}\textemdash which was misattributed to Alhacen from the 14th century until 1967 (\hyperlink{sabra-67}{Sabra, 1967}). I~have noticed that Risner's summarizing headings in Alhacen's work are also sometimes misattributed to Alhacen himself (e.g. in \hyperlink{shapiro-1990}{Shapiro, 1990}, p.\,169, n.\,51, citing \hyperlink{risner-1572}{Risner, 1572}, p.\,129, \S{}8).} It~was Risner's edition that brought the works of Alhacen and Witelo to the attention of Benedetti and Kepler.\footnote{\,\small\hyperlink{goulding-18}{Goulding, 2018}, pp.\,498\textit{n},\,504; \hyperlink{lindberg-81}{Lindberg, 1981}, p.\,185; \hyperlink{smith-2017}{Smith, 2017}, pp.\,322.} And according to Smith, it is Risner's edition that we should blame for changing the spelling of \textit{Alhacen} to \textit{Alhazen} and adding Latin endings thereto.\footnote{\,\small\hyperlink{smith-2001}{Smith, 2001}, p.\,xxi (in the Introduction).}

Witelo, in the second of two postulates (``\textit{petitiones}'') in Book~5 of his \textit{Perspectiva}, says that the location of the object-point with respect to any mirror is taken along the cathetus. He uses this postulate only to establish the cathetus rule in Prop.\,36: ``In~any type of mirror, any visible point is seen on the cathetus of its incidence.'' For the image must be seen according to the aforesaid location of the object-point, or else it will not be seen ``through the mode of image'' (\textit{per modum imaginis}), presumably meaning ``as the image of an object'' and not, e.g., as an independent apparition.\footnote{\,\small\hyperlink{risner-1572}{Risner, 1572}, part~2 (\textit{Vitellonis Opticae}), pp.\,190,\,207, cited by \hyperlink{goulding-18}{Goulding, 2018}, p.\,506; the translations and the interpretation of \textit{per modum imaginis} are Goulding's.}

Kepler rejects Witelo's logic: ``First, I~say that he does not do well to argue from the location of the thing seen to the location of the image, that~is, out of fear that the image might cease to exist if the image should not correspond to the object in position. And indeed, in this way he would easily overturn all of catoptrics. For many things of this sort are different in the image than in the object. Next, for my part, I~do not understand the postulate which he repeats from the beginning of the book,'' except, says~Kepler, for the hint given by Alhacen (\hyperlink{alhacen-obj-img}{above}) in claiming that the state of natural things is in accordance with the situation of their principles.\footnote{\,\small\hyperlink{kepler-donahue-00}{Kepler/Donahue, 2000}, pp.\,74--5.}

In his next proposition, Witelo repeats the cathetus rule for any type of mirror, and tries to prove it by claiming that the image of each point of an extended object must be on the cathetus in order to reproduce the size and shape of the object. But this argument is applicable only to a \textit{plane} mirror, and the resulting geometric transformation of the object is not the only one that would preserve size and shape; e.g., the geometric reflection could be combined with a translation.\footnote{\,\small\hyperlink{goulding-18}{Goulding, 2018}, p.\,507.}

On the cathetus rule for \textit{refraction}, Witelo faithfully recites Alhacen's argument concerning the components of motion. ``It~is hard to see the connection,'' says Kepler, ``and even if you admit it, a mathematical deduction of what was proposed to be proved will not be forthcoming.''{\kern.5bp}\footnote{\,\small\hyperlink{kepler-donahue-00}{Kepler/Donahue, 2000}, p.\,75.}{\tiny\,} Worse, Euclid's ``axiom'' reappears, adapted for refraction. As Kepler reports:
{\small\begin{quote}
To Alhazen's opinion, Witelo appends the view that we had noted above as irrelevant and false in Euclid. He says, ``If on the surface of a transparent body a point upon which there falls a perpendicular from the seen object, happens to be hidden by the interposition of something opaque between the seen object and the point, the object will not be seen.'' I~say that this is false. For provided that the point be free, from which the ray from the seen object to the eye is refracted, the image of the radiating object in the depth will perforce be seen.\footnote{\,\small\hyperlink{kepler-donahue-00}{Kepler/Donahue, 2000}, p.\,76. The statement in quotation marks is italicized in the original Latin edition (\hyperlink{kepler-1604}{Kepler, 1604}, p.\,59), where it is a paraphrase rather than quote from Witelo; \textit{cf}.~\hyperlink{risner-1572}{Risner, 1572}, part~2 (\textit{Vitellonis Opticae}), p.\,415.}
\end{quote}}\noindent
Thus Witelo, after striving through 400 dense pages to improve on Alhacen, regresses 15~centuries in one sentence for a last-ditch defense of the cathetus rule.

\subsection{Benedetti: Binocularism reconsidered}

Kepler has not been alone in his dissatisfaction with the ancient rule. In~a letter to Kepler, written in late 1604 as a critique of the \textit{Paralipomena}, the physician Johannes Brengger proposes a modified rule, which amounts to replacing a reflective surface by its \textit{tangent plane} at the point of reflection before applying the standard cathetus rule.\footnote{\,\small\hyperlink{goulding-18}{Goulding, 2018}, pp.\,533--5.} A~more sophisticated, independent modification is found in the optical writings of \href{https://en.wikipedia.org/wiki/Simon Stevin}{Simon Stevin}, published in 1605.\footnote{\,\small Part of his \textit{Mathematical Memoirs}, first published in Dutch, then translated into Latin by Snell in 1608, and into French (more selectively) in 1608 and again in 1634 (\hyperlink{goulding-18}{Goulding, 2018}, p.\,535; \hyperlink{dijksterhuis-55}{Dijksterhuis, 1955}, pp.\,10,\,30--32, works XIa,\,XIb,\,XIII).} Stevin's rule is a sort of binocular version of Brengger's: each eye sees a ``true'' image in the place given by Brengger; but, in~a curved mirror, convergence of the lines of sight might give the \textit{illusion} of a single image in a third location.\footnote{\,\small\hyperlink{goulding-18}{Goulding, 2018}, pp.\,536--43.} Of~course, what Stevin calls an illusion is what modern readers would regard as the true location of the binocular image.

A more rigorous critic than Brengger and Stevin, and a forerunner of Kepler, is \href{https://en.wikipedia.org/wiki/Giambattista Benedetti}{Giovanni Battista Benedetti}. His \textit{Book of Various Mathematical and Physical Speculations} (Turin, 1585) contains five treatises followed by a miscellany of letters. One of the letters, addressed to a certain Conradus Terl, recognizes the role of the retina in vision, and in so doing may have anticipated \href{https://en.wikipedia.org/wiki/Felix Platter}{Felix~Platter}, although Platter was first to publish.\footnote{\,\small\hyperlink{benedetti-1585}{Benedetti, 1585}, pp.\,296--7; \hyperlink{goulding-18}{Goulding, 2018}, p.\,512.} Of interest here, however, is the series of eight undated letters headed ``on the reflections of rays'' and addressed to ``the most excellent philosopher \href{https://en.wikipedia.org/wiki/Francesco Vimercato}{Francesco Vimercato}''\!,\footnote{\,\small\hyperlink{benedetti-1585}{Benedetti, 1585}, pp.\,331--47.} which, according to \href{https://en.wikipedia.org/wiki/Robert D. Goulding}{Robert Goulding}, were probably written in the early 1570s.\footnote{\,\small\hyperlink{goulding-18}{Goulding, 2018}, pp.\,512--13.}

The first letter of the series gives several examples showing that Hero's principle of least distance does not necessarily apply to a \textit{concave} mirror. In~the first example,\footnote{\,\small\hyperlink{goulding-18}{Goulding, 2018}, pp.\,513--14.} Benedetti shows that if we have a concave spherical mirror, with the object-point \textit{n} and the observation point \textit{q} on the spherical surface (extended if necessary), and seek a reflection point \textit{b} opposite the chord \textit{qn}, the position of \textit{b} is that which \textit{maximizes} the path length,\footnote{\,\small{}Provided that the two legs of the path\textemdash from the object-point to the reflection point, and from the latter to the observation point\textemdash are constrained to be straight; if they are allowed to be curved, the path length is never a local maximum, because it can always be increased via the arc lengths of the legs (cf.~\hyperlink{born-wolf-02}{Born \&~Wolf, 2002}, p.\,137\textit{n}). Concerning the \href{https://en.wikipedia.org/wiki/Fermat's principle}{Hero/Fermat principle}, Goulding (\hyperlink{goulding-18}{2018}, pp.\,513--14) makes two errors in passing. First, in~his footnote~52, he fails to note that a refracted path may be a path of \textit{maximum} time (again subject to the constraint that the legs are permissible ray paths) if the surface of the denser medium is sufficiently convex (consider, e.g., the refracted path through a small glass bead in the middle of the line of sight). Second, in~his footnote~53, referring to the concave spherical mirror, the length of the reflected path ``through the unlabeled end of the diameter \textit{bc}'' is not, as he claims, the ``very shortest'' from \textit{q} to~\textit{n}; as the proposed point of reflection approaches \textit{q} or~\textit{n}, the path length approaches the length of the chord \textit{qn}, which is clearly shorter than the path via any other point on the sphere.} contrary to Hero's teleological principle. Hence Benedetti prefers a \textit{mechanistic} explanation of the law of reflection, which he offers in the third letter of the series.\footnote{\,\small\hyperlink{goulding-18}{Goulding, 2018}, pp.\,514--15, citing \hyperlink{benedetti-1585}{Benedetti, 1585}, p.\,335.} That explanation is unconvincing by modern standards, but sets a fruitful precedent: in the same (third) letter, Benedetti goes on to seek a similarly mechanistic explanation of the cathetus rule\textemdash assuming the use of \textit{two} eyes.

Benedetti is not the first optician to consider \href{https://en.wikipedia.org/wiki/Binocular vision}{binocular vision}; Ptolemy, Alhacen, and Witelo have all confronted it.\footnote{\,\small\hyperlink{goulding-18}{Goulding, 2018}, pp.\,507--9.} But, whereas his predecessors have treated it as a problem\textemdash how to avoid seeing double\textemdash Benedetti treats it as an opportunity: how to perceive depth. Like Alhacen, he understands that if an object-point is to be seen singly and most distinctly, the axes of the two eyes must converge on that point; but, unlike Alhacen, he explicitly associates this convergence of the visual axes with the \textit{distance} at which an object is seen singly, and he recognizes it as the mechanism of distance perception. Idiosyncratically, he adds that the distance is still perceived by looking with \textit{one} eye, because (he~claims) the object is still seen best when the axis of the other eye passes through it.\footnote{\,\small\hyperlink{benedetti-1585}{Benedetti, 1585}, pp.\,335--6; \hyperlink{goulding-18}{Goulding, 2018}, pp.\,515--16.}

Armed with this new understanding of binocular vision, Benedetti considers the reflection of an object-point in a plane mirror, viewed with both eyes. Alhacen has used the cathetus rule to locate the image seen by each eye separately, and concluded that the two images coincide so that ``there will only be one image\ldots\,and it will lie at the same place as it would if it were viewed by only one eye.''{\kern.5bp}\footnote{\,\small\hyperlink{smith-2006}{Smith, 2006}, pp.\,401--3 (Prop.\,4), with notes on pp.\,492--3, and diagrams on p.\,221 (other~volume); \textit{cf}.~\hyperlink{goulding-18}{Goulding, 2018}, pp.\,509--12 (with~diagrams).}{\tiny\,} Benedetti inverts this reasoning: because the two lines of sight, produced through the mirror, intersect the cathetus at the same point, they intersect \textit{each~other} at that point, which is therefore the image\textemdash and on the cathetus. Thus, for the special case of reflection in a plane mirror, Benedetti gives the \textit{first valid proof of the cathetus rule} for lines of sight other than the cathetus itself.\footnote{\,\small\hyperlink{benedetti-1585}{Benedetti, 1585}, p.\,336; \hyperlink{goulding-18}{Goulding, 2018}, pp.\,516--18.}

For a convex spherical mirror,\footnote{\,\small\hyperlink{smith-2006}{Smith, 2006}, pp.\,431--2, with notes on pp.\,499--500, and diagrams on~p.\,240 (other~volume), pars.\,2.217--18.} and (more~tersely) for a concave spherical mirror,\footnote{\,\small\hyperlink{smith-2006}{Smith, 2006}, p.\,475.} Alhacen again relies on the cathetus rule to show that each eye sees the image-point at the same location, provided that the eyes are placed symmetrically about a plane containing the cathetus.\footnote{\,\small{}A statement on binocular perception of images is found at the end of Alhacen's discussion of each mirror shape, with the unexplained exception of the convex cylinder (\hyperlink{goulding-18}{Goulding, 2018}, pp.\,511--12). For a convex conical mirror, Alhacen says that ``the same form and the same location for the form is perceived by each eye\ldots; sometimes they share precisely the same location, sometimes their locations overlap, and sometimes they are separated, but only a little bit'' (\hyperlink{smith-2006}{Smith, 2006}, p.\,446), where this ``little bit'' is apparently small enough to allow ``a~single image according to sense-deduction'' (\hyperlink{smith-2006}{Smith, 2006}, p.\,431). For a concave cylindrical mirror, he baldly asserts that ``when both eyes are looking, one image will actually form two, but they will abut or overlap, so they will appear single'' (\hyperlink{smith-2006}{Smith, 2006}, p.\,481). And he gives a similar statement on what happens when a second eye is opened to each of the images formed by a concave conical mirror (\hyperlink{smith-2006}{Smith, 2006}, p.\,485).} In~the \textit{concave} case, for which Alhacen does not even offer a diagram, Benedetti gives a detailed original argument, which again avoids using the cathetus rule as a premise. Supposing at first that the object-point and both eyes are \textit{on} the reflecting sphere, Benedetti shows that both (produced) lines of sight must intersect the cathetus. But only if the points of reflection are equidistant from the object-point do the intersections coincide, in which case there is a single image-point on the cathetus; otherwise, he says, the two eyes see separate images. We can see that the same reasoning applies if the eyes are moved forward, closer to the cathetus. But, as Benedetti notes, if they cross to the other side of the cathetus the object-point will be seen double and blurred (``\textit{confus\`{e}}''), wherever the points of reflection may be.\footnote{\,\small\hyperlink{benedetti-1585}{Benedetti, 1585}, pp.\,337--9; \hyperlink{goulding-18}{Goulding, 2018}, pp.\,518--20.} Moreover, he says, if the two eyes are in the same plane of reflection (confusingly called the \textit{surface} of reflection), then
{\small\begin{quote}
the place of the image will not be on the cathetus of incidence, but outside it, because the intersection of the visual axes will not be on the cathetus but outside it\textemdash and in that intersection there takes place the vision of only one image, something that the ancients did not notice.\footnote{\,\small\hyperlink{goulding-18}{Goulding, 2018}, p.\,521, quoting \hyperlink{benedetti-1585}{Benedetti, 1585}, p.\,339.}
\end{quote}}\noindent
\hypertarget{sixth}{Thus Benedetti ends the third letter by asserting a \textit{counterexample to the cathetus rule}. He does not give a proof here. In~the sixth letter, however,} he shows that a spherical burning mirror with an object-point beyond the center of curvature does not give a single focal point on the cathetus, and concludes:
{\small\begin{quote}
Whence it follows that the convergence of reflected rays from a concave spherical mirror is not at one and the same point on the cathetus of incidence, when they are reflected from points not equidistant from the same cathetus. From this reasoning it may also be seen that what I wrote to you in the third letter is true, namely that whenever the visual axes or reflected rays are in one and the same plane of reflection, then the image of the object will in no way be seen on the cathetus of incidence in a concave spherical mirror.\footnote{\,\small\hyperlink{benedetti-1585}{Benedetti, 1585}, p.\,343; \textit{cf}.~\hyperlink{goulding-18}{Goulding, 2018}, p.\,521.}
\end{quote}}\noindent
Indeed the violation of the cathetus rule in the third letter involves points of reflection that are not equidistant from the cathetus. Concerning the violation~in the sixth letter, Benedetti apparently reasons that if the reflected rays in a common plane of reflection intersect the cathetus at different points, they must intersect \textit{each other} at points \textit{off} the cathetus, as asserted in the third letter.

In the seventh letter (the last that deals with specular reflection), Benedetti gives another counterexample and another salvage, both for a \textit{convex} spherical mirror. For the counterexample, he considers two rays from the same object-point in the same plane of reflection, and shows that if the reflected rays, when produced, intersect each other on the cathetus, then they cannot both satisfy the law of reflection.\footnote{\,\small\hyperlink{benedetti-1585}{Benedetti, 1585}, pp.\,343--4, summarized in \hyperlink{goulding-18}{Goulding, 2018}, pp.\,523--5.} For the salvage, he takes an object-point \textit{b}, from which the foot of the cathetus is \textit{g}, and shows that if a ray from \textit{b} is reflected with sufficiently glancing incidence at a point \textit{q}, the produced reflected ray intersects the cathetus \textit{bg} in the air \textit{outside} the sphere.\footnote{\,\small\hyperlink{goulding-18}{Goulding, 2018}, p.\,525 \& Fig.\,15.} He concludes that
{\small\begin{quote}
if the reflected rays from the object \textit{b} come to both pupils from two points of such a mirror, as distant from point \textit{g} as \textit{q} is, then the common point of convergence of the visual axes will be on the cathetus\ldots\,where the image will appear for the reasons given above, so that this can happen not only with concave, but also with convex mirrors.\footnote{\,\small\hyperlink{benedetti-1585}{Benedetti, 1585}, p.\,344.}
\end{quote}}\noindent
The salvage does not depend on the point of convergence being outside the sphere. It~depends only on the axial symmetry about the cathetus, which implies that each produced reflected ray intersects the cathetus \textit{somewhere}, and that if the points of reflection are equidistant from the foot of the cathetus, so are the points of intersection.\footnote{\,\small{}Goulding (\hyperlink{goulding-18}{2018}, p.\,526) explains Benedetti's conclusion thus: ``from his analysis of the concave mirror he extrapolated the general principle that any image location predicted by the traditional theory could be saved by the binocular theory, if the eyes were symmetrically placed on either side of the older theory's plane of reflection''\!. I~should add that the symmetry of the surface needs to be axial about the cathetus, and that the lines of sight need to be related by a rotation about the cathetus. If the symmetry were merely bilateral about ``the older theory's plane of reflection''\!, it would guarantee only that the image is in that plane\textemdash not that it is necessarily on the cathetus.}

\subsection{Kepler: Generalized lines of sight}

So the first disproof-and-salvage of the cathetus rule, with the first explicit counterexamples, is due to Benedetti. But here we have heard from Kepler first, because it is to him that we owe the first rebuttals of traditional \textit{arguments} for the rule. In the third chapter of his \textit{Paralipomena}, having disposed of these arguments, Kepler introduces a series of propositions of his own, ``\textit{in order to make evident the true cause of the place of the image}, ignorance of which is a disgraceful stain in a most beautiful science''\!.\footnote{\,\small\hyperlink{kepler-donahue-00}{Kepler/Donahue, 2000}, p.\,76.} For Kepler, as for his predecessors, an image is essentially an illusion:
{\small\begin{quote}
\textit{The Optical writers say it is an image, when the object itself is indeed perceived along with its colors and the parts of its figure, but in a position not its own, and occasionally endowed with quantities not its own, and with an inappropriate ratio of parts of its figure.} Briefly, an image is the vision of some object conjoined with an error of the faculties contributing to the sense of vision. Thus, the image is practically nothing in itself, and should rather be called imagination.\footnote{\,\small\hyperlink{kepler-donahue-00}{Kepler/Donahue, 2000}, p.\,77 (Definition\,1); \textit{cf}.~Malet \hyperlink{malet-1990k}{1990k}, p.\,6.}
\end{quote}}\noindent
But what is the location of this illusory thing? In Proposition~8, Kepler eventually informs us that the distance of the image from the eye(s) is judged by triangulation, ``as~is more amply discussed below concerning \href{https://en.wikipedia.org/wiki/Parallax}{parallaxes}''\!, with a baseline given by the distance between the eyes, or motion of the head, by which ``a~single eye stands in for two that are far apart''\!, or, at worst, the breadth of the pupil, as elaborated in Propositions 9 and 14.\footnote{\,\small\hyperlink{kepler-donahue-00}{Kepler/Donahue, 2000}, pp.\,79--83.}$^,$\footnote{\,\small{}The reference to parallaxes is, I~submit, an admission that a small angle of convergence between the eyes may be judged with the aid of background objects rather than by any innate ability to sense the angle.} Thus he follows Benedetti in referring to triangulation, but goes beyond Benedetti by \textit{allowing baselines other than those given by binocular vision}.

Also in Proposition~9, we read that Nature intended the edges of the eyelids, and the line connecting the eyes, to be in the plane of the horizon in order to maximize the baseline for triangulation within that plane. For that reason, according to Proposition~10, when you look at an object-point via a convex mirror or ``the flat surface of denser media,'' you try to position your eyes so that the two lines of sight meet the surface at equal angles. If this condition is not met, says Kepler (again somewhat cryptically), the two lines of sight generally fail to intersect, so that you see two images, unless you strain your eyes so as to look along skew lines.\footnote{\,\small\hyperlink{kepler-donahue-00}{Kepler/Donahue, 2000}, pp.\,80--81.} (Recall that Benedetti has noted the double vision for asymmetric placement of the eyes, but only for reflection, and only for a \textit{concave} mirror.\footnote{\,\small\hyperlink{goulding-18}{Goulding, 2018}, pp.\,519--20.})

For cases that meet the ``equal angles'' condition, Kepler salvages the cathetus rule.\footnote{\,\small\hyperlink{kepler-donahue-00}{Kepler/Donahue, 2000}, pp.\,83--6.} In Definition~2, he introduces the plane of reflection or refraction (again confusingly called the \textit{surface} of reflection or refraction), which earlier writers have defined as the plane containing the observation point (``center of vision''), the object-point, and the point of reflection/refraction. This plane is perpendicular to the reflecting or refracting surface (Prop.\,16). Now let an object-point be viewed by both eyes via a plane or spherical reflecting or refracting surface (\textbf{Prop.\,17}). For each eye, there is an point of reflection or refraction, and a line of sight (``visual ray'') through that point. The image-point, if one exists for the given positions of the eyes, is the point where these lines of sight meet, which must be on the line of intersection of the respective planes of reflection/refraction (since these contain the lines of sight). These planes contain the object-point and are perpendicular to the surface at the respective points of reflection/refraction, and hence, by the symmetry, contain the cathetus, which is therefore their line of intersection, which (as already established) contains the image-point. Thus ``\textit{all the images of the seen object will be on the perpendicular from the object to the surface, whether refracting or reflecting; and this will happen to such an extent that the distance of the points of the seen object is grasped in the manner described, whether by the two eyes, or by the diameter of the breadth of one eye}.''{\kern.5bp}\footnote{\,\small\hyperlink{kepler-donahue-00}{Kepler/Donahue, 2000}, p.\,86; Kepler's emphasis.}{\tiny\,} And it is grasped in that manner by the two eyes if the two lines of sight make equal angles with the surface (Prop.\,10).

Goulding initially describes Kepler's Prop.\,17 as a ``rapid proof to show that the image seen in a plane mirror would lie on the visible object's cathetus''\!, this proof being ``identical to Benedetti's'' except in ``only two ways'': first, Kepler does not repeat Benedetti's claim that monocular depth-perception involves the alignment of the other eye; and second, Kepler extends the argument to plane refraction. But, as Goulding adds on the same page, ``Kepler intended this argument to apply to any reflective or refractive surface of any shape,'' subject to appropriate symmetry in the placement of the eye(s).\footnote{\,\small\hyperlink{goulding-18}{Goulding, 2018}, p.\,529.} Indeed Kepler himself, in~Prop.\,17, implicitly allows the surface to be spherical,\footnote{\,\small\hyperlink{kepler-donahue-00}{Kepler/Donahue, 2000}, p.\,86 (after the italics). But I~concede that the allowance is only implicit\textemdash which may explain Goulding's incongruous conclusion that Kepler, unlike Benedetti, ``did not provide a proof'' for non-plane mirrors (\hyperlink{goulding-18}{Goulding, 2018}, p.\,531).} but does not say whether it is convex or concave; his reasoning depends solely on axial symmetry about a well-defined cathetus and is otherwise indifferent to the shape of the surface or whether it is reflective or refractive.

\hypertarget{new-salvage}{Here I should mention a case, not mentioned by Benedetti or Kepler, in~which the cathetus rule holds although the ``equal angles'' condition does not.} Recall that \hyperlink{takahashi-defense}{Takahashi defends Euclid} by noting that if you try to look along the cathetus at the reflection of an extended object, your line of sight is blocked. Now this problem does not arise with refraction. Accordingly, consider a smooth refracting surface with the object-point on one side and your eyes on the other, with one eye (the ``first'') on the cathetus, so that the line of sight produced from the first eye through the surface \textit{is} the cathetus. If the surface and media are axially symmetrical about the cathetus, or otherwise bilaterally symmetrical about the plane of the object-point and both eyes, then, by that symmetry, the line of sight produced from the \textit{second} eye through the surface intersects the cathetus. And the point of intersection is the binocular image-point.

Kepler gives his first counterexample to the cathetus rule in Prop.\,18.\footnote{\,\small\hyperlink{kepler-donahue-00}{Kepler/Donahue, 2000}, pp.\,86--8; \textit{cf}.~\hyperlink{goulding-18}{Goulding, 2018}, pp.\,530--31.} Unlike Benedetti, he does not consider a concave mirror in this connection. For a \textit{convex} spherical mirror, like Benedetti, he considers two rays from the same object-point in the same plane of reflection. But, whereas Benedetti supposes that the two (produced) reflected rays meet on the cathetus, and shows that they cannot both satisfy the law of reflection, Kepler supposes the law of reflection and shows by a purely geometric contradiction argument that the (produced) reflected rays meet on the observer's side of the cathetus. Indeed, as he shows more simply, the point at which they meet moves outside the sphere as we approach grazing incidence. He concludes that the cathetus rule is not universally true, ``unless this restriction also be added, that the sense of vision be so located with respect to the mirror as nature shows''{\kern.5bp}\footnote{\,\small\hyperlink{kepler-donahue-00}{Kepler/Donahue, 2000}, p.\,88.}\textemdash that~is, unless the lines of sight make equal angles with the surface.\footnote{\,\small\textit{Cf}.~\hyperlink{darrigol-12}{Darrigol, 2012}, pp.\,27,\,74\textit{n}.} But, he adds, the departure from the cathetus is imperceptible if only one eye is used, because the lines of sight are so close together.

Kepler's theory of image location, including his disproof-and-salvage of the cathetus rule, was thought to be novel until 2018, when Benedetti's partial priority was revealed by Goulding. Kepler himself presents his theory as revolutionary, without citing Benedetti's \textit{Speculations}. Had he known this work, says Goulding, ``such an omission would have been out of character for the usually scrupulous Kepler.''{\kern.5bp}\footnote{\,\small\hyperlink{goulding-18}{Goulding, 2018}, p.\,531.}{\tiny\,} On that score, I~can easily believe that Benedetti and Kepler independently thought of proving the cathetus rule for a plane mirror by inverting Alhacen's binocular argument, because (pardon the anecdote) \hyperlink{ingredients}{so~did~I}, before I~knew that Alhacen had introduced a second eye or a second line of sight. I~can even believe that Benedetti and Kepler (unlike~me) independently thought of supporting their argument by citing the same proposition XI.19 of Euclid's \textit{Elements}, because mathematicians of bygone centuries (unlike~me) knew their Euclid and cited him slavishly. Like Benedetti, Kepler gives the counterexample of the convex mirror with the two eyes in the same plane of reflection; but Goulding concedes that Kepler's treatment is ``more concise and elegant''\!, and I~further submit that it gives more information. Like Benedetti, Kepler rejects Hero's least-distance explanation of the law of reflection (propagated through Alhacen and Witelo), but for different reasons: the variation of the path length is negligible for reflections of stars in ponds, and the argument fails completely for refraction, supporting Kepler's claim that ``these operations are not those of a form that acts deliberately or keeps a goal in mind, but of matter bound to its geometrical necessities.''{\kern.5bp}\footnote{\,\small\hyperlink{kepler-donahue-00}{Kepler/Donahue, 2000}, p.\,84; \textit{cf}.~\hyperlink{goulding-18}{Goulding, 2018}, pp.\,528--9,\,531.}{\tiny\,} There is a letter in which Kepler expresses a high opinion of Benedetti's mathematics\textemdash an opinion which, according to Goulding, he could hardly have formed from works other than the \textit{Speculations}.\footnote{\,\small\hyperlink{goulding-18}{Goulding, 2018}, p.\,531.} But if we accept that assessment, the evidence is still leaky because the letter dates from 16~Nov.\,1606, two years after the \textit{Paralipomena}. On that inconclusive note, I~abandon this subplot and return to Kepler's~treatise.

\hypertarget{first-counterex-refr}{In Proposition~19 of the third chapter, Kepler gives the first counterexample to the cathetus rule for \textit{refraction}.} He considers a plane refracting surface, with the object-point in the denser medium and the two eyes in a common plane of refraction in the rarer medium, and shows that for sufficiently oblique incidence, the image departs from the cathetus toward the observer. He does this without knowing the exact law of refraction, by first supposing that the angle of deviation is the same for the two angles of incidence, and then showing that the departure from the cathetus is greater if, as in fact, a more oblique incidence causes a greater deviation.\footnote{\,\small\hyperlink{kepler-donahue-00}{Kepler/Donahue, 2000}, pp.\,88--9; \hyperlink{goulding-18}{Goulding, 2018}, p.\,530 \&~Fig.\,17. In~the redrawings of Kepler's diagram by Donahue and Goulding, the incidence is not sufficiently oblique: to support the argument, point \textit{D} should be to the left of the (vertical) cathetus from \textit{E}; compare the original in \hyperlink{kepler-1604}{Kepler, 1604}, p.\,73.}$^,$\footnote{\,\small{}In the degenerate case in which one eye is on the cathetus, the binocular image is also on the cathetus; see \hyperlink{new-salvage}{above}.}

Ending Kepler's third chapter, in Prop.\,20, is the \textit{reductio ad absurdum} that begins the present paper: the cathetus rule implies that we can move (e.g.)~a reflected image by deforming the reflective surface in the vicinity of the cathetus while preserving it in the vicinity of the point(s) of reflection\textemdash whereas in fact, as Kepler says, ``it~makes no difference to the place of the image, what sort of mirror surface is placed opposite the object, since the proportions of image formation are all taken from that part of the mirror upon which are the two points of reflection of light to the two eyes.''{\kern.5bp}\footnote{\,\small\hyperlink{kepler-donahue-00}{Kepler/Donahue, 2000}, p.\,90.}$^,$\footnote{\,\small{}The supporting example (\hyperlink{kepler-donahue-00}{Kepler/Donahue, 2000}, pp.\,90--91), in~which Kepler seems to have invented what we now call the \href{https://en.wikipedia.org/wiki/Osculating circle}{osculating circle}, is more sophisticated than it needs to be.}

The imprecision of the distance of the image as judged by \textit{one} eye becomes crucial in the fifth chapter of the same work, where Kepler considers a distant object seen through a glass sphere filled with water. He admits that if the eyes are sufficiently far behind the sphere, the image is seen in the air when viewed stereoscopically with two eyes,\footnote{\,\small\hyperlink{kepler-donahue-00}{Kepler/Donahue, 2000}, pp.\,191, 192 (Prop.\,1). In modern terms, of~course, this image is \textit{real}.} but is seen on the facing surface of the sphere when viewed with one eye,\footnote{\,\small\hyperlink{kepler-donahue-00}{Kepler/Donahue, 2000}, pp.\,194 (Prop.\,6), 208--9 (Prop.\,17).} and may be seen in two places on that surface if both eyes are trained on the surface.\footnote{\,\small\hyperlink{kepler-donahue-00}{Kepler/Donahue, 2000}, p.\,195, Prop.\,7; \textit{cf}.~\hyperlink{malet-1990k}{Malet, 1990k}, pp.\,10--12 \&~Fig.\,5.} As \href{https://en.wikipedia.org/wiki/Alan E. Shapiro}{Alan E.~{\tiny\!}Shapiro} points out, this case shows that the \textit{perceived} image and the \textit{geometrical} image (Shapiro's terms) of the same object-point may have different locations, the former image being located by a pair of rays, and the latter by a \textit{pencil} of rays (Kepler's term).\footnote{\,\small\hyperlink{shapiro-1990}{Shapiro, 1990}, pp.\,106--7, 124--5 \&~n.\,58.}

Later in the fifth chapter, Kepler considers refraction of parallel rays by a spherical surface. For deviations less than 10~degrees, using the approximation that the deviations are proportional to the angles of incidence, he shows that the refracted rays cut the axis at very nearly the same point.\footnote{\,\small\hyperlink{kepler-donahue-00}{Kepler/Donahue, 2000}, pp.\,205--6 (Prop.\,15).} Then he introduces what we call the real image, which he calls a \textit{picture} (Latin \textit{pictura}), and which, by his definition, seems to require a screen upon which it appears:
{\small\begin{quote}
\textit{Since hitherto an Image has been a Being of the reason, now let the figures of objects that really exist on paper or upon another surface be called pictures}.\footnote{\,\small\hyperlink{kepler-donahue-00}{Kepler/Donahue, 2000}, p.\,210 (``Definition''); \textit{cf}.~\hyperlink{malet-1990k}{Malet, 1990k}, p.\,14.}
\end{quote}}\noindent
The subsequent Propositions 20 \& 23, which concern the picture projected by a water-filled glass sphere, imply that in order to make an intelligible picture, the rays originating from one point on the object need not converge exactly to one point in the picture; \textit{near}-convergence is enough. In both cases, the ``last intersection''\textemdash \,that~is, the limit of the intersection of the refracted ray with the axis, as the incident ray deviates less and less from the axis\textemdash is recognized as an image, implying that an image need not be perfectly stigmatic.\footnote{\,\small\hyperlink{kepler-donahue-00}{Kepler/Donahue, 2000}, pp.\,211--13.}

But, as noted by \href{https://en.wikipedia.org/wiki/Antoni Malet}{Antoni Malet}\textemdash against the view of previous 20th-century scholars\textemdash it is not at all clear that Kepler regards a geometrical image as acting on the eye in the same way as an object. In~his \textit{Paralipomena}\footnote{\,\small\hyperlink{kepler-donahue-00}{Kepler/Donahue, 2000}, p.\,192 (Prop.\,1).} and in his \textit{Dioptrice} of 1611, in~cases where a real image is formed in the air, Kepler conspicuously fails to invoke it in explaining what is seen by the eye(s) without a screen,\footnote{\,\small Malet, \hyperlink{malet-1990k}{1990k}, pp.\,5 (n.\,8), 21--23; \hyperlink{malet-2003}{2003}, pp.\,118--20,\,134; \hyperlink{darrigol-12}{Darrigol, 2012}, p.\,35 (Fig.\,1.19).} although he \textit{does} invoke it in explaining how an upright picture can be subsequently formed on paper through another lens.\footnote{\,\small\hyperlink{malet-2003}{Malet, 2003}, p.\,120 \&~Figure\,7.}

\medskip

\noindent These three points\textemdash
\begin{itemize}\itemsep=-0.4ex
\item[(i)] that the perceived and geometrical images may not coincide,
\item[(ii)] \hypertarget{point-ii}{that the convergence of the rays may be approximate}, and
\item[(iii)] that a geometrical image is not yet declared to be visually equivalent to an object
\end{itemize}
\textemdash are revisited later in the century.

\subsection{Harriot and Snell: Forgotten triumph}

Meanwhile the cathetus rule has been ironically implicated in an exasperating turn of events: the unpublished rediscoveries of the law of refraction by \href{https://en.wikipedia.org/wiki/Thomas Harriot}{Thomas Harriot} in~1601, and \href{https://en.wikipedia.org/wiki/Willebrord Snellius}{Willebrord Snell} in~1621.

It seems that Harriot immersed a vertical circular disk in water up to its center, sighted object-points on the rim using the center as the point of refraction, and noted that the image-points, \textit{when located according to the cathetus rule}, lay on a smaller \textit{circle} coaxial with the disk.\footnote{\,\small\hyperlink{lohne-59}{Lohne, 1959}, pp.\,116--7; \hyperlink{schuster-00}{Schuster, 2000}, pp.\,274--5.} It~follows that the distances from the point of refraction to the object- and image-points are in a fixed ratio (the ratio of the radii of the outer and inner circles), so~that the \textit{cosecants} of the angles of incidence and refraction are in the same ratio, and their sines are in the inverse ratio.\footnote{\,\small\hyperlink{goulding-22}{Goulding, 2022}, p.\,183.}

Snell's surviving statement of the law begins by saying that the true ray and the apparent ray are in a fixed ratio\textemdash which is true for refraction in a plane surface, if we understand that the ``true~ray'' is measured from the point of refraction to the object-point, and the ``apparent ray'' from the point of refraction to the image-point \textit{as~located by the cathetus rule}. The statement goes on to relate the ray lengths to the cosecants of the angles.\footnote{\,\small\hyperlink{vollgraff-1936}{Vollgraff, 1936}, p.\,720.}

But the later rediscovery by \href{https://en.wikipedia.org/wiki/René Descartes}{Descartes}\textemdash the first discovery of the law of refraction to become public\textemdash is expressed in terms of sines, not cosecants or ray lengths, and shows no other apparent influence by the cathetus rule.\footnote{\,\small{}The case of Descartes' co-worker \href{https://en.wikipedia.org/wiki/Claude Mydorge}{Claude Mydorge} is less clear. Schuster (\hyperlink{schuster-00}{2000}, pp.\,271,\,275--6) is impressed by the similarity between Harriot's diagram and Mydorge's, for which Goulding (\hyperlink{goulding-22}{2022}, pp.\,191--6) offers a different explanation.}

\subsection{Mersenne, Roberval, Gregory: Images redefined}

\href{https://en.wikipedia.org/wiki/Marin Mersenne}{Marin Mersenne}, in his posthumous \textit{L'Optique, et la Catoptrique} (1651) edited by \href{https://en.wikipedia.org/wiki/Gilles de Roberval}{Gilles Personne de Roberval}, distinguishes between two images of the same object: the ``interior or sensible image''\!, which is formed on the retina, and the ``exterior or apparent'' image, ``which our fantasy represents to us some place outside far or near from us, as~if the object itself were in that place, from which it sends its rays to us to form the interior image\ldots''{\kern.5bp}\footnote{\,\small Quoted and translated by Shapiro, \hyperlink{shapiro-2008}{2008}, p.\,311.}

Roberval, in his editorial contribution, refers~to
{\small\begin{quote}
the apparent place of the exterior image of a point of an object in all manners of vision\textemdash direct, reflected, or refracted\textemdash both for one eye alone as for two, being the point where the rays that fall on the eyes concur really or potentially (French: \textit{en~puissance}) immediately before the eyes\ldots\footnote{\,\small \hyperlink{shapiro-2008}{Shapiro, 2008}, p.\,295.}
\end{quote}}\noindent
In modern terms, of course, a point of ``potential'' concurrence is a \textit{virtual} image. Roberval also allows the rays to be \textit{very nearly} concurrent,\footnote{\,\small \hyperlink{shapiro-2008}{Shapiro, 2008}, p.\,294.} in agreement with Kepler on \hyperlink{point-ii}{point~(ii)} above; but the unification of the perceived and geometrical images in the ``exterior'' image, and their visual equivalence to an object (``as~if the object itself were in that place\ldots''), differ with Kepler on points (i) and~(iii).

\href{https://en.wikipedia.org/wiki/James Gregory (mathematician)}{James Gregory}'s \textit{Optica Promota} of 1663\textemdash chiefly known for the invention of \href{https://en.wikipedia.org/wiki/Gregorian telescope}{a~reflecting telescope} (in the Epilogue), the independent rediscovery of the law of refraction (Proposition~4),\footnote{\,\small Discussed at length by Malet (\hyperlink{malet-1990g}{1990g}).} and the preface belatedly acknowledging Descartes' prior publication of this law, of which Gregory was unaware until he went to press\footnote{\,\small{}Gregory's ignorance of Descartes' priority is one of several pieces of evidence suggesting that the propagation of the law of refraction was slow for the first twenty years after its publication by Descartes in 1637; see \hyperlink{dijksterhuis-04}{Dijksterhuis, 2004}, p.\,173.}\textemdash is of interest here for its definition of an image, which is apparently independent of Mersenne and Roberval,\footnote{\,\small \hyperlink{shapiro-2008}{Shapiro, 2008}, p.\,295.} and which parts with Kepler on all of points (i) to~(iii) above. According to Gregory,
{\small\begin{quote}
\textit{An image is a similitude of a radiating body, arising from the divergence or convergence of the rays belonging to individual points of the radiating body, from individual points or to individual points of a single surface.}\footnote{\,\small ``\textit{Imago est similitudo materi\ae\ radiantis, orta ex divergenti\^{a}, vel convergenti\^{a} radiorum, singulorum materi\ae\ radiantis punctorum, a punctis singulis, vel ad puncta singula unius superficiei.}'' \textemdash\,\hyperlink{gregory-1663}{Gregory, 1663}, p.\,1 (Definition~9).}
\end{quote}}\noindent
This definition, like Roberval's, allows no distinction between perceived and geometrical image-points and applies to both binocular and monocular viewing,\footnote{\,\small\hyperlink{gregory-bruce-06}{Gregory/Bruce, 2006}, Props.\,28,\,29,\,36; \hyperlink{shapiro-1990}{Shapiro, 1990}, pp.\,128--30.} and attributes the ``similitude'' to the defining feature of a geometrical image: the convergence or divergence of the rays. But, unlike Roberval (and~Kepler), Gregory does not allow the point of divergence or convergence to be an approximation or limiting case; concerning the image of an object-point \textit{B} seen by reflection, Gregory writes:
{\small\begin{quote}
From the points of the pupil [\textit{A}], draw through the points of reflection all the lines of reflection, in whose concourse \textit{L} (provided they concur) will be the apparent place of the image of the point \textit{B}. If, however, they do not concur in one point, no distinct and fixed place of the image of the visible point \textit{B} will exist.\footnote{\,\small End of Prop.\,36, as translated by Shapiro (\hyperlink{shapiro-1990}{1990}, p.\,129).}
\end{quote}}\noindent
Although the diagram supporting this statement shows the point of concourse as being behind the mirror (giving a virtual image), the wording is equally applicable if the point is in front (giving a real image). Moreover, the initial statement of the problem indicates that the solution should be equally applicable to refraction,\footnote{\,\small\hyperlink{gregory-bruce-06}{Gregory/Bruce, 2006}, Prop.\,36.} which it is. And indeed the initial definition is applicable to both real and virtual images, and to both reflection and refraction. Gregory's insistence on exact concurrence may look like a loss of generality, but is understandable in view of his coverage of the exact imaging properties of conic sections in reflection and refraction.\footnote{\,\small\textit{Cf}.~Malet, \hyperlink{malet-1990g}{1990g}.}

Mersenne, Roberval, and Gregory have not addressed the cathetus rule directly; but their refinement of the concept of the image will be pivotal in a high-profile case.

\subsection{Tacquet: Affirmation and exception}

According to Malet:
{\small\begin{quote}
By the late sixteenth century it was a well-known fact that [distant] things perceived through convex lenses appear inverted or upright according to the distance from the eye to the lens. Empirical accounts of the properties of convex lenses, such as \href{https://en.wikipedia.org/wiki/William Bourne (mathematician)}{William Bourne}'s `Treatise on the properties and qualities of glasses for optical purposes' (1585),\footnote{\,\small Printed in \hyperlink{halliwell-1839}{Halliwell, 1839}, pp.\,32--47. ``1585'' is \href{https://en.wikipedia.org/wiki/Albert Van Helden}{Van Helden}'s dating of the treatise, whereas \href{https://en.wikipedia.org/wiki/Sven Dupré}{Dupr\'{e}} dates it to 1579/80 (\hyperlink{dupre-10}{Dupr\'{e}, 2010}, pp.\,137--8).} did not fail to mention that\, (1)~when the eye is removed from the lens beyond the `burnynge beame', or focus, all [distant] things seen through the lens appear inverted, and\, (2)~when the eye lies between the burning focus and the lens all things seen through the lens appear upright and enlarged, and the more so the closer the eye to the focus.\footnote{\,\small\hyperlink{malet-2003}{Malet, 2003}, p.\,116; ``[distant]'' is my addition, for context. Compare Bourne, chapters~VI to~VIII, in \hyperlink{halliwell-1839}{Halliwell, 1839}, pp.\,42--4. On the contrivance mentioned at the end of chap.\,VI and elaborated in chap.\,IX, see \hyperlink{dupre-10}{Dupr\'{e}, 2010}.}
\end{quote}}\noindent
Here we are chiefly interested in Malet's point~(2), under which we should also note that when the eye reaches the focus, as Bourne says, ``yow shall discerne nothinge thorowe the glasse: But like a myst, or water''\!.\footnote{\,\small\hyperlink{halliwell-1839}{Halliwell, 1839}, p.\,44.}

Kepler explains point (2) in his \textit{Dioptrice}. He shows that when an object-point is viewed through a convex lens at such a distance that the refracted rays converge toward another point, with the eye between that point and the lens, the object is seen upright (Proposition~70) and blurred (``\textit{confusa}''), the more blurred as the eye is further from the lens, since the convergence is greater (Prop.\,71), and most blurred when the eye reaches the point of convergence (Prop.\,74). Moreover the image is magnified (Prop.\,80), and the more so as the eye recedes from the lens toward the point of convergence (Prop.\,82).\footnote{\,\small\hyperlink{kepler-1859}{Kepler (1859)}, pp.\,542--7. The location of these passages was assisted by Darrigol (\hyperlink{darrigol-12}{2012}, pp.\,34--5), Malet (\hyperlink{malet-2010}{2010}, pp.\,283--6), Shapiro (\hyperlink{shapiro-1990}{1990}, p.\,160 \&~n.\,184), and \textit{translate.google.com}. On Kepler's explanation of Prop.\,82, see \hyperlink{malet-2003}{Malet, 2003}, p.\,114 \&~Figure~4. Props.\,80\,\&\,82 are used in Kepler's subsequent explanation of the magnifying power of a Dutch telescope; see Malet, \hyperlink{malet-2003}{2003} at p.\,122, or \hyperlink{malet-2010}{2010} at p.\,286.}

Gregory, in the following passage, confirms the blur but is indifferent to whether the convergence is caused by a lens or a mirror:
{\small\begin{quote}
\textit{Corollary 4.}

\ldots\,[I]f the rays from one point converge toward another point behind the eye [\textit{post~oculum}], no place can be assigned to this point except (if we will) behind the eye at the concourse of the rays: hence the image formed of such points may conveniently be called an image behind the eye.

\hypertarget{gregory-prop-30}{\textbf{Prop.\,30. Theorem.}}

\textit{With the rays from one point converging toward a point situated behind the eye, it is impossible to make distinct vision.}

For every eye is so constructed as to see distinctly either remote [points], which radiate as if in parallel, or near ones, which send out diverging rays; but in no eye is the retina distinctly painted by the converging rays (which originate from artifice and not from nature), because the crystalline humor\footnote{\,\small{}That~is, the lens.} gathers [\textit{congregat}] these rays into a point in the vitreous humor, and sends them disgregated to the retina, from which disgregation arises blurred vision\textemdash as shown by Kepler.\footnote{\,\small Translated from \hyperlink{gregory-1663}{Gregory, 1663}, p.\,41, and in some places differing from \hyperlink{gregory-bruce-06}{Gregory/Bruce, 2006}.}
\end{quote}}\noindent
This ``image behind the eye'' is what we would now call a \textbf{virtual object} presented to the eye. Although there is no mention of the cathetus rule here, the \textit{mirror version of the same experiment}\textemdash in which rays converge from a concave mirror toward a point behind the eye\textemdash is the only case in which the cathetus rule is \textit{not} upheld by \href{https://en.wikipedia.org/wiki/André Tacquet}{Andr\'{e} Tacquet S.J.}~in his \textit{Catoptrica Tribus Libris Exposita} (Catoptrics explained in three books), posthumously published in 1669. At the end of Book~1, Tacquet says of the cathetus rule:
{\small\begin{quote}
This theorem is the most fruitful of all of catoptrics, whereby nearly all the phenomena of plane and convex mirrors are demonstrated, as will become evident from all of book two and book three. Consequently, its truth is in turn extraordinarily established: for it cannot be false, since it agrees wonderfully with all phenomena without exception.\footnote{\,\small\hyperlink{tacquet-1669}{Tacquet, 1669}, p.\,223, quoted in translation by Shapiro (\hyperlink{shapiro-1990}{1990}, p.\,144).}
\end{quote}}\noindent
But he immediately adds:
{\small\begin{quote}
\textit{Whether and when this proposition has a place with concave mirrors will be plain from what is to be said in Book~3}.\footnote{\,\small\hyperlink{tacquet-1669}{Tacquet, 1669}, p.\,223, italics in the Latin.}
\end{quote}}\noindent
And in Book~3, just before Proposition~22,\footnote{\,\small\hyperlink{tacquet-1669}{Tacquet, 1669}, p.\,256.} he warns that ``in concave ones we postulate this only for the moment, until the extent of its truth becomes apparent.'' In~Props.\,29\,\&\,30,\footnote{\,\small\hyperlink{tacquet-1669}{Tacquet, 1669}, p.\,259.} he comes to the experiment just mentioned, in~which the eye intercepts converging rays from a concave mirror. Here the cathetus rule locates the image \textit{behind} the eye\textemdash in agreement with Gregory's terminology\textemdash whereas the mind inevitably construes any visible image as being \textit{in~front} of the eye, leading Tacquet to conclude:
{\small\begin{quote}
\textit{Therefore Alhazen, Witello, and other opticians following them err in considering that just as in plane and convex mirrors so in concave ones the image never appears outside the intersection of the reflected ray with the cathetus of incidence.}
\end{quote}}\noindent
The quote is translated by Shapiro,\footnote{\,\small\hyperlink{shapiro-1990}{Shapiro, 1990}, p.\,172, n.\,107; italics in the Latin.} who further reports that as late as 1735, \href{https://en.wikipedia.org/wiki/Samuel Clarke}{Samuel~Clarke} faulted Tacquet for making even that exception to the cathetus rule,\footnote{\,\small\hyperlink{rohault-clarke-1735}{Rohault/Clarke, 1735}, p.\,278\textit{n}.} while \href{https://en.wikipedia.org/wiki/Christian Wolff (philosopher)}{Christian Wolff} upheld the rule for two eyes provided that they were not in the same plane of incidence.\footnote{\,\small\hyperlink{shapiro-1990}{Shapiro, 1990}, p.\,172, n.\,108.} In~allowing the eyes to be asymmetrically placed in different planes of incidence, Wolff's proviso is too permissive\textemdash as Benedetti and Kepler knew.

\subsection{Barrow ``destroys'' the doctrine}

The Rev.~{\tiny\!}\href{https://en.wikipedia.org/wiki/Isaac Barrow}{Isaac Barrow}, inaugural \href{https://en.wikipedia.org/wiki/Lucasian Professor of Mathematics}{Lucasian Professor} at Cambridge, in~the first of his \textit{Lectiones~XVIII} (Eighteen Lectures) published in 1669, defines images thus:
{\small\begin{quote}
\ldots\,Images are clearly nothing other than light from objects so reflected or refracted that it is again collected in one place and in such a situation as it had when it flowed from the original object and proceeded in a direct path to the eye; whereby it happens that images represent objects similarly but as if they were located elsewhere.\footnote{\,\small\textit{Lectiones} I:5 (\hyperlink{barrow-1669}{Barrow, 1669}, p.\,4, quoted in translation by Shapiro, \hyperlink{shapiro-1990}{1990}, p.\,107).}
\end{quote}}\noindent
In the third lecture he reprises the idea:
{\small\begin{quote}
Indeed by the term \textit{image}, I~understand nothing but the place from which a number of rays (as many as suffice to affect vision) seem to diverge or spread in the same manner as when they are diffused by primary objects.\footnote{\,\small\textit{Lectiones} III:16 (\hyperlink{barrow-1669}{Barrow, 1669}, p.\,30), cited (not translated) by Shapiro (\hyperlink{shapiro-1990}{1990}, p.\,166, n.\,6); my italics.}
\end{quote}}\noindent
As Shapiro explains,\footnote{\,\small\hyperlink{shapiro-1990}{Shapiro, 1990}, pp.\,106--7 (\&~n.\,5), 124--5,\,165.} Barrow's principle of image location, which was rightly linked to him in the 18th century, was wrongly credited to Kepler in the 20th. In~fact Barrow agrees with Roberval: he~follows Roberval and Gregory, against Kepler, by~strictly equating the perceived and geometrical images, and by recognizing the manner in which an image imitates an object; but, as we shall see, he follows Kepler and Roberval, against Gregory, by not requiring an image to be strictly stigmatic.

The case of the eye intercepting converging rays, whether from a convex lens as in Kepler's example, or from a concave mirror as in Tacquet's, is known as the \textbf{Barrovian~case}\footnote{\,\small\hyperlink{shapiro-1990}{Shapiro, 1990}, pp.\,144,\,159--65.} because it is taken up by Barrow\textemdash citing Tacquet but, strangely, not Kepler in this connection\textemdash at the end of his lectures; the relevant passage has been translated from the Latin by \href{https://en.wikipedia.org/wiki/George Berkeley}{Berkeley} and, independently, by Clarke.\footnote{\,\small\hyperlink{berkeley-1901}{Berkeley (1901)}, pp.\,137--40; \hyperlink{rohault-clarke-1735}{Rohault/Clarke, 1735}, pp.\,260--61\textit{n}.\, Fay's recent translation of all eighteen lectures (\hyperlink{barrow-fay-87}{Barrow/Fay, 1987}) is apparently out of print.} Here Barrow notes that because diverging rays appear to come from a finite distance, and parallel rays from an infinite distance, converging rays ought to appear to come from beyond infinity,\footnote{\,\small\hyperlink{berkeley-1901}{Berkeley (1901)}, p.\,138.} whereas in fact, in the case in question, the image may seem closer than the object, and certainly seems to come closer as the rays become more convergent\footnote{\,\small Shapiro (\hyperlink{shapiro-1990}{1990}, p.\,160, line~6) erroneously has \textit{divergence} instead of \textit{convergence}.}\textemdash that~is, as the eye recedes toward the point of convergence\textemdash until ``the object appearing extremely near begins to vanish into mere confusion.''{\kern.5bp}\footnote{\,\small\hyperlink{berkeley-1901}{Berkeley (1901)}, p.\,139.}{\tiny\,} Indeed the image seems to come closer because (as mentioned by Kepler but not Barrow) the magnification increases, and because (as mentioned by neither, but easily observed) the direction of the image becomes more sensitive to sideways movement of the eye\textemdash although the apparent movement of the image is the wrong way for an image in \textit{front} of the eye. As Barrow notes, the looming of the image offends not only ``our~Notion'' (his principle of image location), but also ``that antient and common one'' (the cathetus rule):
{\small\begin{quote}
It seems so much to overthrow that antient and common one, which is more a-kin to ours than any other, that the learned Tacquett was forced by it to renounce that Principle, (upon which alone, almost all his Catoptricks depend) as uncertain, and not to be depended upon, whereby be overthrew his own Doctrine.\footnote{\,\small\hyperlink{rohault-clarke-1735}{Rohault/Clarke, 1735}, p.\,261\textit{n}.}
\end{quote}}\noindent
After this caricature of Tacquet's position, Barrow immediately concedes:
{\small\begin{quote}
Which, nevertheless, I~do not believe he would have done, had he but considered the whole matter more thoroughly, and examined the difficulty to the bottom.\footnote{\,\small\hyperlink{berkeley-1901}{Berkeley (1901)}, p.\,139; this statement is elided in Clarke's translation.}
\end{quote}}\noindent
The concession is startling\textemdash the more so for its want of explanation\textemdash in that it seems to imply that Tacquet's purported counterexample to the cathetus rule is \textit{not} a counterexample. That indeed is the position subsequently taken by Clarke, who argues that the cathetus rule is not in play, because the reflected rays, being intercepted by the eye, do not meet the cathetus.\footnote{\,\small\hyperlink{rohault-clarke-1735}{Rohault/Clarke, 1735}, p.\,278\textit{n}.} In~his commentary on the Barrovian passage, Clarke explains the apparent closeness of the image by noting that\, (i)~if the eye is sufficiently close to the point of convergence, we cannot simultaneously train both eyes on the object-point through the glass (however large it may be), and with only one eye the judgment of distance is inferior and influenced by the proximity of the glass, and\, (ii)~as the eye recedes, the increasing magnification (and brightness, in the case of a luminous object) makes the image seem to come closer.\footnote{\,\small\hyperlink{rohault-clarke-1735}{Rohault/Clarke, 1735}, p.\,262\textit{n}.} Berkeley's explanation,\footnote{\,\small\hyperlink{berkeley-1901}{Berkeley (1901)}, pp.\,140--43 (\S\S\,31,\,35--6); \textit{cf}.~\hyperlink{cardona-gutierrez-20}{Cardona \&~Guti\'{e}rrez, 2020}.} although earlier, is more modern, noting that the convergence of rays via a lens or mirror is not the only reason why an object may appear blurred; another is that the object is too \textit{close}!

A late twist in the story of the Barrovian case\textemdash presumably unknown to all the players from Bourne in the 16th century to Clarke in the 18th\textemdash is that the concave-mirror version, including the application of the cathetus rule, is discussed in Ptolemy's \textit{Optics}.\footnote{\,\small Experiment~IV.1, translated in \hyperlink{smith-1996}{Smith, 1996}, pp.\,194--5, with further commentary in \hyperlink{smith-2017}{Smith, 2017}, pp.\,104--7.} For a given position of the eye and a given point of reflection, Ptolemy marks three object positions for which the cathetus rule will place the image respectively at the eye, behind the eye, and nowhere (or, as we would say, at infinity), and states the range of object positions for which the rule places the image behind the mirror. For the case in which the rule would place the image behind the eye, he claims that the object seems to be in front of the mirror (contrary to the rule) because the visual faculty is biased toward the surface from which the reflection comes. Similarly, when the rule places the image at infinity or at the eye, Ptolemy says it is perceived to be \textit{on the mirror}. Later, for a single spherical surface, Ptolemy gives what would amount to a refractive version of the experiment, if it were described in the same detail.\footnote{\,\small Theorem~V.9, translated in \hyperlink{smith-1996}{Smith, 1996}, p.\,252, with commentary in \hyperlink{smith-2017}{Smith, 2017}, p.\,119 \&~figure\,3.15.} Less likely to have escaped notice is the related example given by Alhacen,\footnote{\,\small\hyperlink{smith-2006}{Smith, 2006}, p.\,451 (par.\,2.331) and figure~5.2.34b on p.\,254 (other~volume); \hyperlink{risner-1572}{Risner, 1572}, p.\,162, reprised by Witelo at his pp.\,314--5.} and cited by Bacon,\footnote{\,\small\hyperlink{bacon-combach-1614}{Bacon/Combach, 1614}, pp.\,139--40; \hyperlink{bacon-burke-1928}{Bacon/Burke, 1928}, pp.\,553--5; \hyperlink{smith-2017}{Smith, 2017}, p.\,268.} for which the cathetus rule places one of the images behind the eye. Here Alhacen does not comment on the evident impossibility, whereas Bacon, like Ptolemy, blames the limits of vision:
{\small\begin{quote}
But in all these diversities of appearances the image is never truly apprehended unless its place is beyond the mirror, or between the sight and the mirror; hence what appears in the center of the eye or behind the head is not perceived there. For vision is not born to apprehend the positions of forms unless they are in front of it.\footnote{\,\small\hyperlink{bacon-combach-1614}{Bacon/Combach, 1614}, p.\,140; \textit{cf}.~\hyperlink{bacon-burke-1928}{Bacon/Burke, 1928}, p.\,555.}
\end{quote}}\noindent

In the Barrovian case, in the words of Barrow's definitions of an image, the point toward which the rays converge is neither ``light\ldots\,again collected in one place''\!, because the light never gets there, nor a place from which rays ``seem to diverge''\!, because they \textit{con}verge. (That~is, in modern terms, it~is neither a real image nor a virtual image.) Therefore, according to Barrow's criteria, it should not be the perceived image. But what should be? Barrow does not have an answer that passes the test of experiment. So we are forced to admit that in the Barrovian case, as in all the other cases surveyed by Tacquet (if~he is to be believed), the ancient cathetus rule does no worse than Barrow's post-Keplerian principle of image location.\footnote{\,\small{}In modern terms, the point toward which the rays converge in the Barrovian case is a virtual object presented to the front surface of the eye, which refracts the rays toward a nearer point, which in turn becomes a virtual object presented to the interface between the cornea and the aqueous humor, and so on, until a real image is formed in front of the retina. From this image the rays diverge again to form a blurred picture \textit{on} the retina (as~Gregory notes in his \hyperlink{gregory-prop-30}{Prop.\,30}, quoted above). What is presented to the observer's retina is thus easily explained and uncontroversial. What the observer makes of it is another matter: ``Insofar as I~can determine''\!, says Shapiro (\hyperlink{shapiro-1990}{1990}, p.\,178, n.\,206), ``there is still no generally accepted explanation for the `Barrovian case.'\,''}

However, Barrow's principle manifestly does better than ``that antient and common one'' in explaining another case: the location of the image seen by refraction in a plane surface, which Barrow determines by some inspired pre-calculus geometry and ``the most recently given law or hypothesis of refraction (discovered by the illustrious Descartes, but now, I~believe, embraced by most of the better Opticians\ldots)''\!.\footnote{\,\small Translated from \hyperlink{barrow-1669}{Barrow, 1669} (introduction); \textit{cf}.~\hyperlink{shapiro-1990}{Shapiro, 1990}, p.\,113.}

\begin{figure}[p]\centering
\includegraphics[width=3.6in]{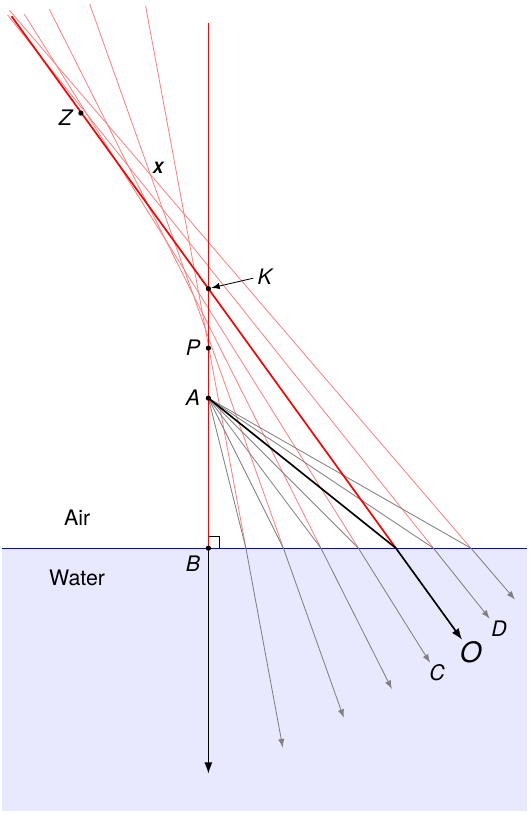}
\caption{Isaac Barrow's location of the tangential image~\textit{Z} of an object-point~\textit{A} seen by an observer at~\textit{O} due to refraction. The tangential image is the point of tangency between the refracted ray produced back from~\textit{O}, and the \textit{caustic} (common tangent curve) of all the other produced refracted rays from the same object-point in the same plane of refraction. Point~\textit{K} is the image location given by the old cathetus rule; it~lies on the cathetus~\textit{AB}. Point~\textit{P}, where the caustic meets the cathetus, is the \textit{paraxial} image, i.e.~{\tiny\!}the image of~\textit{A} seen by an observer on the cathetus, below~\textit{B}. (Diagram by the author, after Barrow.)}
\label{f-Barrow}
\end{figure}

Given an object-point \textit{A} in the rarer medium, another point \textit{X} in that medium, and the constraint that the (produced) refracted ray must pass through~\textit{X}, Barrow seeks the refracted ray. He~finds that there are two solutions which merge under a certain condition, under which he renames \textit{X} as \textit{Z} and supposes that the eye (at~\textit{O}) looks along the refracted ray, which thereby becomes what he calls the ``principal ray'' (\textit{ZO}~in our \hyperref[f-Barrow]{Figure~{\tiny\!}4}), i.e.~{\tiny\!}the ray through the center of the eye.\footnote{\,\small\hyperlink{shapiro-1990}{Shapiro, 1990}, pp.\,130,\,132--3.} He then argues that \textit{Z} is where the eye sees the image, because if we take two neighboring refracted rays from the same object-point~\textit{A} in the same plane of refraction, one on each side of the principal ray (e.g., the rays passing through \textit{C} and~\textit{D}), and produce them back through the interface, they intersect the principal ray~\textit{ZO} on opposite sides of the point~\textit{Z}. And this point, as~he has found, is \textit{beyond the cathetus} with respect to the eye.\footnote{\,\small\hyperlink{shapiro-1990}{Shapiro, 1990}, pp.\,133--4.} Whereas ``Alhazen and most of the multitude of opticians after him'' would place the image at~\textit{K}, i.e.~{\tiny\!}at the intersection of the produced principal ray~\textit{ZO} and the cathetus~\textit{AB}, Barrow notes that only one ray from~\textit{A} (namely~\textit{AO}) is produced back through~\textit{K} unless the eye is on the cathetus,\footnote{\,\small\hyperlink{shapiro-1990}{Shapiro, 1990}, p.\,134 \&~n.\,86, quoting \textit{Lectiones}~V, \S{}21, incorrectly numbered 20 in the original printing (\hyperlink{barrow-1669}{Barrow, 1669}, pp.\,44--6).} in which case, as he shows in the previous lecture,\footnote{\,\small\hyperlink{shapiro-1990}{Shapiro, 1990}, pp.\,131--2.} all refracted rays that enter the pupil will, when produced back through the interface, intersect the cathetus at nearly the same point (marked~\textit{P} in our \hyperref[f-Barrow]{Figure~{\tiny\!}4}).

If the eye is \textit{off} the cathetus, the image-point~\textit{Z} found by Barrow is what we now call the \textbf{tangential} image\textemdash because it is the point of tangency between the (produced) line of sight and the \textbf{\href{https://en.wikipedia.org/wiki/Caustic (optics)}{caustic}} (common tangent curve) of all the (produced) refracted rays originating from the same object-point in the same plane of refraction.\footnote{\,\small\hyperlink{darrigol-12}{Darrigol, 2012}, pp.\,73--4; \hyperlink{shapiro-1990}{Shapiro, 1990}, pp.\,109,\,139. The term \textit{caustic}\textemdash but not the concept\textemdash was apparently coined in 1690 by \href{https://en.wikipedia.org/wiki/Ehrenfried Walther von Tschirnhaus}{Ehrenfried Walther von Tschirnhaus} (\hyperlink{darrigol-12}{Darrigol, 2012}, pp.\,28,\,74--5; \hyperlink{shapiro-1990}{Shapiro, 1990}, pp.\,157--8 \&~n.\,165).} This tangency explains his procedure: for the given object-point~\textit{A}, there can be two refracted rays produced through the target point~\textit{X} if~\textit{X} is \textit{off} the caustic, but only one if it is \textit{on} the caustic.\footnote{\,\small\textit{Cf}.~\hyperlink{shapiro-1990}{Shapiro, 1990}, p.\,108, Figure~1.}

If, on the contrary, the eye is \textit{on} the cathetus (below~\textit{B}), the image-point found by Barrow is the cusp of the caustic (our point~\textit{P}), which is now known as the \textbf{\href{https://en.wikipedia.org/wiki/Paraxial approximation}{paraxial}} image, and which \textit{satisfies the cathetus rule in the limiting case}.\footnote{\,\small{}Barrow finds the paraxial image before he finds the tangential image. That the former is the limit of the latter follows from the displayed equation on p.\,148 of \hyperlink{shapiro-1990}{Shapiro, 1990}, by letting $i$ and~$r$ approach zero, so that their cosines approach~1, yielding the paraxial equation on p.\,147. These equations are for a spherical surface, but are easily adapted for a plane surface by putting $\rho\!\to\!\infty$.} Barrow refers to the tangential image as the ``relative'' image, which is ``mutable'' and ``less important''\!, and to the paraxial image as the ``absolute'' image, which is ``simple'' and ``principal''\!.\footnote{\,\small\hyperlink{shapiro-1990}{Shapiro, 1990}, pp.\,109,\,136.} In both cases he applies the term ``image'' to a point that \textit{nearly} coincides with all the intersections between rays entering the pupil from the same object-point; in~this he follows Kepler and Roberval, against Gregory.\footnote{\,\small\hyperlink{shapiro-1990}{Shapiro, 1990}, pp.\,128--30; \hyperlink{kepler-donahue-00}{Kepler/Donahue, 2000}, pp.\,211--13 (Props.\,20,\,23); \hyperlink{gregory-bruce-06}{Gregory/Bruce, 2006}, Prop.\,36.}

Nowadays we tend to think of the tangential image in contradistinction to the \textbf{sagittal} image. The~latter, Barrow ignores;\footnote{\,\small\hyperlink{shapiro-1990}{Shapiro, 1990}, p.\,172, n.\,101.} where he says that only one (produced) refracted ray passes through the image-point alleged by the cathetus rule (point~\textit{K}), he~implicitly confines his attention to rays in the same plane of refraction on the same side of the cathetus. It~is left to his successor and former student, \href{https://en.wikipedia.org/wiki/Isaac Newton}{Isaac Newton}, to point out that in consequence of the axial symmetry about the cathetus, a whole cone of refracted rays shares this property, giving a second image-point~(\textit{K}), which is now called the sagittal image, and which \textit{exactly satisfies the cathetus rule}.\footnote{\,\small\hyperlink{newton-anon-1728}{Newton/anon.,\,1728}, pp.\,104--5\,(scholium)\,\&\,Plate\,7; \hyperlink{shapiro-1990}{Shapiro,\,1990}, pp.\,135--6,\,172.} Recall, however, that Newton's observation is partly anticipated by Kepler, who considers two rays in the said cone,\footnote{\,\small\hyperlink{kepler-donahue-00}{Kepler/Donahue, 2000}, pp.\,85--6 (Prop.\,17); \textit{cf}.~\hyperlink{darrigol-12}{Darrigol, 2012}, p.\,74, and \hyperlink{shapiro-1990}{Shapiro, 1990}, p.\,121.} but subsequently ignores the sagittal image.\footnote{\,\small\hyperlink{shapiro-1990}{Shapiro, 1990}, pp.\,123--4.}

Interchanging the dense and rare media, we return to the \hyperlink{first-counterex-refr}{case considered by Kepler} in which (e.g.)~one looks into still water from above, with the eyes in a common plane of refraction.\footnote{\,\small\hyperlink{kepler-donahue-00}{Kepler/Donahue, 2000}, pp.\,88--9 (Prop.\,19).} Here Barrow offers the following ``not inelegant'' experiment, which confirms the proposition of Kepler (not cited) and ``clearly destroys the doctrine of Alhazen and his followers''\!.\footnote{\,\small\hyperlink{shapiro-1990}{Shapiro, 1990}, pp.\,134--5 \&~n.\,89, quoting \textit{Lectiones}~V, \S{}22, incorrectly numbered 21 in the original printing (\hyperlink{barrow-1669}{Barrow, 1669}, p.\,46).} Attach a weight \textit{F} to a string and hang it from a pivot \textit{G}, with \textit{G} above the water's surface and \textit{F} below, adjusting the height and depth so that, when your eyes are level and facing the string, the refracted image of \textit{F} appears just below the reflected image of \textit{G}. With your eyes in this natural attitude, the two images indeed appear aligned with the string and its reflected image\textemdash that~is, on the cathetus. But now tilt your head so that both eyes are in a common plane of reflection/refraction, and the refracted image of \textit{F} has moved toward you, away from the reflected image of \textit{G}\textemdash that~is, away from the cathetus, in defiance of the ancient rule. Seeing is believing.\footnote{\,\small{}Yes, I~\textit{did} try this at home.}

For oblique reflection in a convex spherical mirror, Barrow's ``relative'' image, like Kepler's image with the eyes in a common plane of reflection,\footnote{\,\small\hyperlink{kepler-donahue-00}{Kepler/Donahue, 2000}, pp.\,86--8 (Prop.\,18).} is on the observer's side of the cathetus. Considering the object-point as a general point on an infinitely long line perpendicular to the mirror, Barrow shows that the image of the line is curved and angled to it, whereas the cathetus rule, ``gratuitously assumed and contrary to reason''\!, would have the image in line with the object. But, in an apparent reference to Tacquet\textemdash who claims to have verified experimentally ``a hundred times'' that the image is in line, and backs the claim by appealing to the axial symmetry about the cathetus,\footnote{\,\small\hyperlink{tacquet-1669}{Tacquet, 1669}, p.\,222 (Prop.\,19).} although the line of sight violates that symmetry\textemdash Barrow concedes that the deviation of this image from the cathetus is harder to observe than the deviation of the refracted image in the aforesaid plumb-line experiment, with the eyes in a common plane of refraction: there the reflected image marks the cathetus, and the refracted image is manifestly not on it.\footnote{\,\small\hyperlink{shapiro-1990}{Shapiro, 1990}, pp.\,142--3, quoting Barrow, \textit{Lectiones}~XVI.}

\subsection{Newton and the ``axiom'' of stigmatism}

Newton's salvage of the cathetus rule for the sagittal image, in the case of axial symmetry about the cathetus, is relegated to his posthumously published \textit{Optical Lectures}.\footnote{\,\small\hyperlink{newton-anon-1728}{Newton/anon., 1728}, pp.\,104--5 (scholium) \&~Plate~7; \hyperlink{shapiro-1990}{Shapiro, 1990}, pp.\,135--6.} In~his better-known \textit{Opticks}, the first 19~pages consist of eight definitions followed by eight ``Axioms and their Explications''\!, by which he then claims to have given ``the sum of what hath hitherto been treated of in Opticks'' or at least ``what hath been generally agreed on''\!.\footnote{\,\small\hyperlink{newton-2010}{Newton (2010)}, pp.\,19--20.}

``Despite his grandiose claim,'' says Shapiro,\footnote{\,\small\hyperlink{shapiro-1990}{Shapiro, 1990}, p.\,149.} ``he did do a remarkable job of compressing elementary geometrical optics into nine pages.'' The compression begins with the following ``axiom'' on p.\,10:
{\small\begin{quote}
\textbf{Ax.~VI.}

\textit{Homogeneal Rays which flow from several Points of any Object, and fall perpendicularly or almost perpendicularly on any reflecting or refracting Plane or spherical Surface, shall afterwards diverge from so many other Points, or be parallel to so many other Lines, or converge to so many other Points, either accurately or without any sensible Error. And the same thing will happen, if the Rays be reflected or refracted successively by two or three or more Plane or Spherical Surfaces}.

The Point from which Rays diverge or to which they converge may be called their \textit{Focus}.\,\ldots
\end{quote}}\noindent
In other words, for reflection or refraction by a plane or spherical surface, if the angles of incidence are not too large, the image of the object-point (although the term \textit{image} has not yet been introduced) will be near enough to \textit{stigmatic}, at least for ``homogeneal'' (monochromatic) rays. This axiom leads to four rules, stated without proof, for locating the focus of the rays reflected or refracted by a plane surface (``\textit{Cas}.\,1''), reflected by a spherical surface (``\textit{Cas}.\,2''), refracted by a spherical surface (``\textit{Cas}.\,3''), and refracted by a lens (``\textit{Cas}.\,4''). Here we should emphasize, although Newton does not, that in the first three cases\textemdash those which involve a single surface and a single cathetus\textemdash the stated location of the focus is \textit{on the cathetus}.

In his next ``axiom'' (p.\,14), Newton gives the condition under which a set of foci makes a picture; but, unlike Kepler, he implicitly acknowledges the independent existence of the foci:
{\small\begin{quote}
\textbf{Ax.~VII.}

\textit{Wherever the Rays which come from all the Points of any Object meet again in so many Points after they have been made to converge by Reflection or Refraction, there they will make a Picture of the Object upon any white Body on which they fall}.
\end{quote}}\noindent
Thence he explains the \href{https://en.wikipedia.org/wiki/Camera obscura}{camera obscura}, the eye, long- and short-sightedness, and correcting spectacles.

In the final ``axiom'' of the set (p.\,18), he endorses Barrow's principle of image location without naming Barrow or using the word \textit{image}:
{\small\begin{quote}
\textbf{Ax.~VIII.}

\textit{An Object seen by Reflexion or Refraction, appears in that place from whence the Rays after their last Reflexion or Refraction diverge in falling on the Spectator's Eye}.
\end{quote}}\noindent
For a plane mirror, he explains, if that place of divergence is point \textit{a}, ``these Rays do make the same Picture in the bottom of the Eyes as if they had come from the Object really placed at \textit{a}\ldots'' As further examples he cites a prism with refracted rays diverging from \textit{d}, and a lens with refracted rays diverging from~\textit{q}. Then he abruptly refers to the ``Image of the Object'' at~\textit{q}{\tiny\,} as~having a certain size, and goes on to use the term \textit{image} routinely, without further introduction. But he has implied, immediately after Ax.\,VI, that a place of divergence is a ``focus''\!, allowing us to interpret that ``axiom'' as giving sufficient conditions for the approximate stigmatism of the image.

Now let us consider the implications of stigmatism. For brevity, we shall follow Barrow by using the term \textbf{inflection} to mean either reflection or refraction.\footnote{\,\small\hyperlink{shapiro-1990}{Shapiro, 1990}, pp.\,130,\,136,\,171\,(n.\,78), citing \hyperlink{barrow-1669}{Barrow, 1669}, pp.\,10\,(\S\,11),\,22,\,111.}$^,$\footnote{\,\small{}Not until 1675 was the term \textit{inflection} hijacked for diffraction by \href{https://en.wikipedia.org/wiki/Robert Hooke}{Hooke} and Newton; see \hyperlink{darrigol-12}{Darrigol, 2012}, pp.\,92--3 \&~n.\,29.}

If the image of an object-point in the inflecting surface is \textit{stigmatic}, it is the common point of intersection of all the inflected rays (for a real image), or of all the inflected ray-lines produced back through the surface (for a virtual image); in either case, it is a \textit{point of intersection of all lines of sight} to the object-point via the surface (produced rectilinearly through the surface if necessary). Hence a ray incident along the cathetus, when ``inflected'' (and produced if necessary), passes through the same image-point. But that ray is \textit{undeviated}: it is transmitted without refraction or reflected back along itself, so that the ``inflected'' ray and the resulting line of sight remain on the cathetus. Thus the image-point lies at the intersection of the cathetus and any other line of sight (whether the image is real or virtual). Conversely, if the image-point lies at the intersection of the cathetus and the line of sight, then, if ``the'' image-point is to be consistent, all such lines of sight must intersect the cathetus at the same point, and therefore must intersect each other at that point, which is therefore a stigmatic image. In~short:
\begin{quote}\large
The cathetus rule is equivalent to the proposition that \emph{the image of the object-point is stigmatic within the working aperture, which admits the cathetus}.
\end{quote}
Notice that the derivation of this equivalence \textit{does not depend on any law of reflection or refraction} except that a normally-incident ray is undeviated. Thus the equivalence, whatever its importance or lack thereof, may be rightly assigned a status that the ancients wrongly assigned to the cathetus rule itself: the status of being as fundamental as the laws of reflection and refraction.

In the case of the sagittal image formed by inflection at a surface axially symmetrical about the cathetus, the image is stigmatic within a working aperture consisting of two infinitesimal areas, one containing the foot of the cathetus and the other containing a circle with its axis on the cathetus.

The cathetus admitted by the working aperture may be notional provided that it is unambiguous, so that we cannot move the cathetus without moving the ``\hyperlink{active}{active}'' part of the surface. For example, while the conditions of Newton's ``Ax.\,VI'' do not say that the working aperture admits the cathetus, they do say that the inflecting surface is plane or spherical, which implies that it can be uniquely produced (extended) so as to admit a unique undeviated ray\textemdash the ``notional'' cathetus\textemdash for a given object-point. And under these conditions, according to the ``axiom''\!, the image is stigmatic ``either accurately or without any sensible Error.''

So, after the cathetus rule has been reduced to a peculiarity of the sagittal image and dismissed from the elementary teaching of optics, a proposition implying wider conditions under which the rule holds, ``either accurately or without any sensible Error''\!, is put up as \textit{axiomatic} at the beginning of the introductory treatise by the highest authority on the subject!

In the statements and applications of the cathetus rule by ancient and medieval opticians, the assumption of stigmatism is always unrecognized and sometimes patently absurd. Alhacen's retention of the rule for cylindrical and conical mirrors may be consigned to the absurd category, except in cases of bilateral symmetry about the plane of reflection, for which the working aperture may be reduced to an infinitesimally narrow strip; in those cases the assumption of stigmatism may still be inexact, but is at least not absurd. In~the unrecognized category, but \textit{almost} recognized, are the cases which exploit the axial symmetry to claim that the image-point is on the cathetus although it is viewed from off the cathetus; this reasoning tacitly assumes that the image-point stays put as the line of sight moves off the cathetus, which is true if the various lines of sight have a common intersection. For example, Alhacen, having established that the image of the center of the eye in a convex spherical mirror is on the cathetus, extends the argument to another point on the eye, although that point is seen from off the cathetus;\footnote{\,\small\hyperlink{smith-2006}{Smith, 2006}, pp.\,396--7.} and Tacquet argues from the same symmetry that the image of a rod aligned with the cathetus is likewise aligned with the cathetus, although it is best seen from off the cathetus.\footnote{\,\small\hyperlink{tacquet-1669}{Tacquet, 1669}, p.\,222 (Prop.\,19).} Apparently the first writer to recognize the \textit{necessity} of stigmatism is Benedetti, who, in~his sixth letter to Vimercato (see~\hyperlink{sixth}{above}), introduces the counterexample of the spherical burning mirror by saying ``I~will prove to you that at no point can all the reflected rays meet each other.''{\kern.5bp}\footnote{\,\small\hyperlink{benedetti-1585}{Benedetti, 1585}, p.\,342.}

But in the useful range of cases that satisfy the conditions of Newton's ``Ax.\,VI''\textemdash \,that the surface is plane or spherical, and that the angles of incidence are not too large\textemdash ancient and medieval investigators should indeed have found the cathetus rule to be true ``either accurately or without any sensible Error.'' That range of cases also includes the following:
\begin{itemize}\itemsep=-0.4ex
\item When we look nearly vertically into still water, the departure of the image from the cathetus is imperceptible, as conceded by Kepler,\footnote{\,\small\hyperlink{kepler-donahue-00}{Kepler/Donahue, 2000}, p.\,89, end of Prop.\,19.} confirmed by Barrow,\footnote{\,\small\hyperlink{shapiro-1990}{Shapiro, 1990}, pp.\,131--2; the diagram is upside-down for an air-water surface.} and implied in Newton's ``\textit{Cas}.\,1.''
\item The same applies to looking nearly vertically \textit{out of} the water (also covered by ``\textit{Cas}.\,1''), as shown by Barrow, who also implies that the `absolute' image is the limit of the `relative' (tangential) image as~the eye approaches the cathetus,\footnote{\,\small\hyperlink{shapiro-1990}{Shapiro, 1990}, pp.\,132--4.} which he calls the \textit{axis} or \textit{radiant~axis}.\footnote{\,\small\hyperlink{shapiro-1990}{Shapiro, 1990}, p.\,108.}$^,$\footnote{\,\small{}Not to be confused with what he calls the ``optical~axis''\!, which is synonymous with his ``principal~ray'' and passes through the center of the eye (\hyperlink{shapiro-1990}{Shapiro, 1990}, pp.\,137,\,141,\,171\,n.79).}
\item Parallel incident rays refracted by a spherical surface, with small deviations, cut the axis at nearly the same point, as noted by Kepler,\footnote{\,\small\hyperlink{kepler-donahue-00}{Kepler/Donahue, 2000}, pp.\,205--6 (Prop.\,15).} and by Barrow,\footnote{\,\small\hyperlink{shapiro-1990}{Shapiro, 1990}, p.\,144.} who shows that an object-point at a finite distance gives the same result,\footnote{\,\small\hyperlink{shapiro-1990}{Shapiro, 1990}, p.\,147.} in agreement with Newton's ``\textit{Cas}.\,3.''
\end{itemize}

These are some of the reasons why Newton's ``axiom''\!, together with the case of the sagittal image, have been useful enough to launch the cathetus rule on a second, incognito career.

\section[A cathetus by any other name\ldots\vspace{.3ex}]{A cathetus by any other name\ldots}

\subsection{Anon.}

The equivalence between stigmatism and the cathetus rule is apparent in any diagram that shows a single surface bringing many rays from a single object-point to a focus at a single image-point, with one of the rays perpendicular to the surface. The (actual or assumed) stigmatism of the image is shown by the concurrence of the lines, and the point of concurrence is the point where every refracted or reflected ray (produced if necessary) meets the undeviated perpendicular ray\textemdash the cathetus. Such diagrams are offered in the widely-used text by Jenkins \&~White (\hyperlink{jenkins-white-76}{1976}) on pp.~{\tiny\!}47, 48, 49, and 100 (Fig.\,6B), the first and last being for an object at infinity. In~each of these cases, the authors take the surface to be spherical (so~that the stigmatism is only approximate) and the perpendicular ray is identified only by its passing through the center of curvature.

If the image of an object-point is stigmatic, it is uniquely located by \textit{any two rays} belonging to that object-point, and we might as well choose those rays for convenience. For a single surface, the most obvious convenience is to let one of the rays be the one along the cathetus, so that it is undeviated. The location of the image then becomes a straightforward but unacknowledged application of the cathetus rule. This is how the image-point is located in our \hyperref[f-Forna]{Figure\,1} above. This is how Jenkins \&~White (\hyperlink{jenkins-white-76}{1976}, pp.\,56--7) and \href{https://en.wikipedia.org/wiki/George S. Monk}{Monk} (\hyperlink{monk-63}{1963}, pp.\,8--9) derive the ``Gaussian formula'' relating the object and image distances for a spherical refracting surface\textemdash without explaining that the generality of the angles implies the stigmatism of the image within the accuracy of the formula.

\begin{figure}[t]\centering
\includegraphics[width=2.5in]{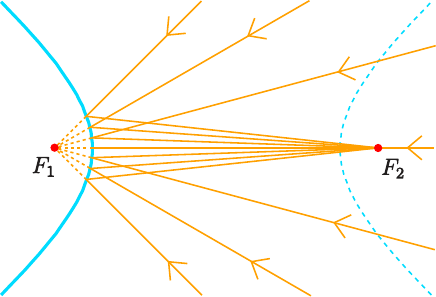}
\caption{Stigmatic image $F_2$ of a virtual object-point~$F_1\:\!$, formed by reflection in a convex hyperboloidal mirror with foci $F_1$ and~$F_2$. Rays initially directed toward~$F_1$ are reflected through~$F_2\:\!$, including the undeviated ray~$F_2\:\!F_1\:\!$, which is the cathetus. Thus the image-point is the intersection of the cathetus with any other reflected ray. (\href{https://commons.wikimedia.org/wiki/File:Hyperbolic_mirror.svg}{Diagram by `Episcophagus'} at \textit{Wikimedia Commons}.)}
\end{figure}

\subsection{Axis}

The convenience of choosing a ray along the cathetus is multiplied if the object-point is on the axis of a system with several coaxial surfaces, so that the axis is perpendicular to all the surfaces. Then the image formed by the first surface is on the axis, which is therefore the cathetus for the second surface, which therefore forms another image on the axis, and so on, so that the axis serves the common cathetus for all the surfaces, and the final image is where the final refracted or reflected ray cuts that common cathetus. Thus Jenkins \&~White (\hyperlink{jenkins-white-76}{1976}) explain how to locate the image of an object-point on the axis of two thin lenses (pp.\,68--9, Fig.\,4I), or of one thick lens (pp.\,78--9, Fig.\,5A),\footnote{\,\small\textit{Cf}.~\hyperlink{hecht-17}{Hecht, 2017}, p.\,167, Fig.\,5.14 (b)~\&~(c).} especially for an object-point at infinity (pp.\,84--5, Fig.\,5G); the intermediate steps need not detain us (yet), except that their purpose is to find where the final refracted ray cuts the axis, because ``the axis itself is considered as the second light ray'' (p.\,69; \textit{cf}.~p.\,79). The beginnings of this approach may be discerned in \href{https://en.wikipedia.org/wiki/Bonaventura Cavalieri}{Bonaventura Cavalieri}'s ``Six Geometrical Exercises'' of 1647.\footnote{\,\small\hyperlink{cavalieri-1647}{Cavalieri, 1647}, p.\,464ff; \hyperlink{shapiro-1990}{Shapiro, 1990}, pp.\,127--8.}

But Barrow calls the cathetus the axis where there is only one surface, axially symmetrical about it.\footnote{\,\small\hyperlink{shapiro-1990}{Shapiro, 1990}, p.\,108.} Jenkins \&~White (\hyperlink{jenkins-white-76}{1976}) do likewise in diagrams showing the focal points of a spherical refracting surface (p.\,46; four cases)\footnote{\,\small Cases (b) and (d) are respectively equivalent to Figs.\,5.11 and 5.10 in Hecht (\hyperlink{hecht-17}{2017}, p.\,165), except that Hecht does not name the ``axis''\!.} and a spherical reflecting surface (p.\,99; two cases)\footnote{\,\small Also in Fig.\,5.61 in Hecht (\hyperlink{hecht-17}{2017}, p.\,196).}; and in those cases where the image is at a finite distance, its assumed stigmatism is seen from the concurrence of the ray-lines, and its location is seen to be consistent with the cathetus rule.

Jenkins \&~White (\hyperlink{jenkins-white-76}{1976}, pp.\,56--7) and Monk (\hyperlink{monk-63}{1963}, pp.\,8--9) even use the word \textit{axis} in their derivations of the ``Gaussian formula''\!, albeit only in the text. Here Jenkins \&~White make ad-hoc approximations from the outset, and Monk does so at the second step. For reasons which will become apparent, we shall now re-derive this formula in a more disciplined manner, introducing assumptions only as they are needed, after showing what can be deduced without them.

\begin{figure}[t]\centering
\includegraphics[width=\textwidth]{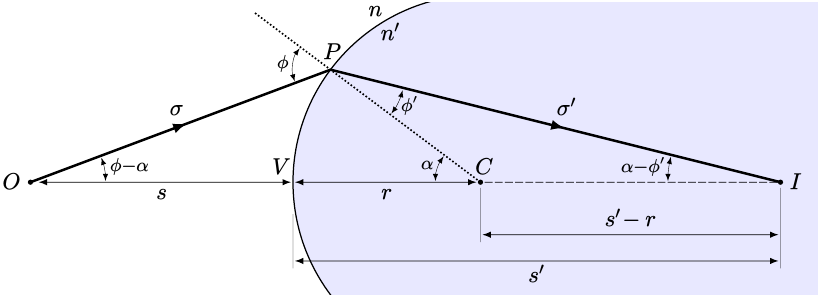}
\caption{Distances and angles for refraction at a spherical surface. (Diagram by the author.)}
\label{f-sagit}
\end{figure}

Let $O$ be an object-point facing a spherical refracting surface (separating two homogeneous isotropic media) whose radius of curvature is $r$ (positive if convex as seen from~$O$) with center~$C$, so that $\mathit{OC}$ is the cathetus (\hyperref[f-sagit]{Figure~{\tiny\!}6}). Let $V$ (for~\textit{vertex}) be the foot of the cathetus, at~a distance $s$ from~$O$.\, Let the point of refraction be~$P$.\, The \textit{axial} symmetry of the interface and media about the cathetus~$\mathit{OC}$ implies a bilateral symmetry about the plane of the cathetus and the incident ray~$\mathit{OP}$, which in turn implies that the refracted ray must remain in that plane.\footnote{\,\small{}Alternatively we can argue that by the bilateral symmetry, the normal to the surface at~$P$ is in the plane of symmetry, which is therefore the plane of the incident ray and the normal, whence, by the law first articulated by Ptolemy, the refracted ray is in that plane. But I~submit that the symmetry is enough, and that the law of Ptolemy follows from it.} So let the point~$I$, at a distance $s'$ from~$V$, be the intersection of the refracted ray and the cathetus (if the refracted ray is parallel to the cathetus, we shall consider $I$ to be at infinity). If angle $\mathit{OCP}$ is called~$\alpha$, then, treating $\alpha$ and~$\phi$ (in~\hyperref[f-sagit]{Figure~{\tiny\!}6}) as exterior angles of triangles, we find that the remote interior angles at $I$ and~$O$ are respectively ${\alpha\text{\textminus}\phi'}$ and ${\phi\text{\textminus}\alpha}$ (as~labeled).

Now it is clear from the symmetry that $s'$ is an \textit{even} function of ${\alpha\text{\textminus}\phi'}$. This, together with the smoothness of the function (apart from the \href{https://en.wikipedia.org/wiki/Removable singularity}{removable singularity} at ${\alpha\text{\textminus}\phi'\text{=}\,0}$), implies that the graph of $s'$ vs.~${\alpha\text{\textminus}\phi'}${\tiny\,} passes through the $s'$ axis with a slope of zero, so that the intersection $I$ is stationary as the observation point (on~$\mathit{PI}$, beyond~$I$) passes through the cathetus~$\mathit{OC}$. For the given object-point~$O$, this stationarity of~$I$ is the limit of the intersection of a refracted ray with the cathetus as ${\alpha\text{\textminus}\phi'\text{\textrightarrow\,0}}$ (as~claimed by Barrow), hence the limit of the intersection of two refracted rays with each other as both approach the cathetus, hence the limit of the tangential image-point as the observation point approaches the cathetus (as shown by Barrow). The limiting position of~$I$, by construction, is on the cathetus, salvaging the cathetus rule as an approximation for small angles; and because the limit is a stationarity, the deviation from the limit, measured along the cathetus, is at worst 2nd-order in the angles (in which case the ray aberration is of 3rd order, as expected). This implies near-stigmatism for sufficiently small angles\textemdash justifying Newton's ``axiom''\!.

All this has been shown from symmetry and smoothness, without relying on the exact law of refraction\textemdash or even the exact sphericity of the surface, provided that it is axially symmetric about the cathetus and sufficiently smooth. But now let us invoke the sphericity with center~$C$, so that the segment~$\mathit{CP}$ (in~\hyperref[f-sagit]{Figure~{\tiny\!}6}) has length~$r$. Let the distances $\mathit{OP}$ and~$\mathit{PI}$ be respectively $\sigma$ and~$\sigma'$ (as~shown). Then, by the \href{https://en.wikipedia.org/wiki/Law of sines}{sine~rule} in triangle~$\mathit{OCP}$, we have
\[\frac{r}{\,\sigma\,} = \frac{\sin(\phi-\alpha)}{\sin{\alpha}}\]
or, after expanding the sine of the difference and simplifying,
\begin{equation}
  \frac{r}{\,\sigma\,} = \sin\phi\,\cot\alpha - \cos\phi \,.
\label{e1}\end{equation}
Similarly, applying the sine rule in triangle~$\mathit{ICP}$ (and noting that the exterior angle has the same sine as its supplementary interior angle), we have
\[\frac{r}{\,\sigma'} = \frac{\sin(\alpha-\phi')}{\sin{\alpha}} \,,\]
i.e.
\begin{equation}
  \frac{r}{\,\sigma'} = \cos\phi' -\, \sin\phi'\cot\alpha \,.
\label{e2}\end{equation}
To eliminate $\alpha$, we multiply~(\ref{e1}) by ${\tfrac{\sin\phi'}{r}}$,\, and~(\ref{e2}) by ${\tfrac{\sin\phi}{r}}$,\, and add the results, obtaining
\begin{equation}
  \frac{\sin\phi'}{\sigma} + \frac{\sin\phi}{\sigma'}
  \,=\, \frac{\sin\phi\,\cos\phi' -\, \cos\phi\,\sin\phi'}{r} \,.
\label{e3}\end{equation}

For the purpose of locating~$I$, let us rearrange~(\ref{e3}) as
\begin{equation}
  \frac{1}{\,\sigma'} \,=\, \frac{\sin(\phi-\phi')}{r\sin\phi}
                            - \frac{\sin\phi'}{\sigma\sin\phi} \,.
\label{e4}\end{equation}
Then, for paraxial rays, the angles $\phi$ and~$\phi'$ are small so that the sines may be approximated by their arguments, and $\sigma$ and~$\sigma'$ may be approximated by $s$ and $s'$ respectively, the fractional errors being 2nd-order in the angles. Thus we have
\begin{equation}
  \frac{1}{\,s'} \approx \frac{\,\phi-\phi'}{r\phi} - \frac{\,\phi'}{s\phi} \,.
\label{e5}\end{equation}
As $\mathit{CP}$ is the radius of the spherical interface (\hyperref[f-sagit]{Figure~{\tiny\!}6}), it is the normal to the interface at~$P$, whence $\phi$ and~$\phi'$ are the angles of incidence and refraction. Kepler did not know the exact law of refraction (although he had corresponded with Harriot, who did\footnote{\,\small\hyperlink{lohne-59}{Lohne, 1959}; \hyperlink{shirley-51}{Shirley, 1951}.}); but he was satisfied that for small angles, the ratios ${\tfrac{\phi\text{\textminus}\phi'}{\phi}}$ and ${\tfrac{\,\phi'}{\phi}}$ are approximately constant,\footnote{\,\small That he was aware of this fact as early as 1604 is shown in \hyperlink{kepler-donahue-00}{Kepler/Donahue, 2000}, pp.\,124, 127-9 (Prop.\,8), \&~205--6 (Prop.\,15)\textemdash although he made greater use of it in his \textit{Dioptrice} of 1611, where it is stated up-front as ``VII.~Axioma'' [\hyperlink{kepler-1859}{Kepler (1859)}, p.\,529]. \textit{Cf}.~\hyperlink{darrigol-12}{Darrigol, 2012}, pp.\,34--5; \hyperlink{dijksterhuis-99}{Dijksterhuis, 1999}, p.\,29; \hyperlink{malet-2003}{Malet, 2003}, p.\,109; \hyperlink{shapiro-1990}{Shapiro, 1990}, pp.\,126--7.} in which case, by~(\ref{e5}), for given $r$ and $s$, the length $s'$ is approximately constant. The same conclusion applies to \textit{reflection}, for which we put ${\phi'\text{=\,\textminus}\phi}$ in~(\ref{e5}) and write $-s'$ for~$s'$ (that~is, change the positive direction of~$s'$), obtaining
\begin{equation}
  \frac{\,1\,}{s} + \frac{1}{\,s'} \approx -\frac{2}{\,r\,} \,.
\label{e6}\end{equation}
Barrow first published this result.\footnote{\,\small Expressed as an equation by Shapiro (\hyperlink{shapiro-1990}{1990}, p.\,140), and matching \hyperlink{jenkins-white-76}{Jenkins \&~White, 1976}, p.\,103, Eq.\,(6b). On the priority of Barrow (vs.\,Huygens), see Shapiro, p.\,128, and \hyperlink{dijksterhuis-99}{Dijksterhuis, 1999}, pp.\,39,\,86.}

Having seen what can be done \textit{without} the exact law of refraction, let us now invoke~it: if $n$ and~$n'$ denote the refractive indices of the two media (\hyperref[f-sagit]{Figure~{\tiny\!}6}), then the ratio ${\tfrac{n}{\sin\phi'}}$ is the same as ${\tfrac{n'}{\sin\phi}}$.\, Multiplying the exact equation~(\ref{e3}) by this ratio, in the first form for terms in ${\sin\phi'}$ and the second for terms in ${\sin\phi}$, we get
\begin{equation}
  \frac{n}{\,\sigma\,} + \frac{\,n'}{\,\sigma'}
  = \frac{n'\cos\phi' -\, n\cos\phi}{r} \,.
\label{e7}\end{equation}
For paracathetal/paraxial rays, the cosines may be replaced by 1\, while $\sigma$ and~$\sigma'$ may be replaced by $s$ and~$s'$ (the fractional errors again being 2nd-order in the angles), to obtain
\begin{equation}
  \frac{n}{\,s\,} + \frac{\,n'}{\,s'} \approx \frac{n'\! - n}{r} \,,
\label{e8}\end{equation}
which is well known as the \textbf{Gaussian~formula} for a spherical refracting surface,\footnote{\,\small\hyperlink{jenkins-white-76}{Jenkins \&~White, 1976}, pp.\,48,\,56.} although Barrow again gives an equivalent result.\footnote{\,\small\hyperlink{shapiro-1990}{Shapiro, 1990}, p.\,147 (for refractive indices 1 and~$n$).} For \textit{reflection}, we put ${\phi'\text{=\,\textminus}\phi}${\tiny\,} and ${n'\text{=\,\textminus}n}${\tiny\,} in (\ref{e7}) and~(\ref{e8}){\tiny\,} and change the positive directions of $\sigma'$ and~$s'$, obtaining
\begin{equation}
  \frac{1}{\,\sigma\,} + \frac{1}{\,\sigma'} = -\frac{2\cos\phi}{r}
\label{e9}\end{equation}
for the exact result, and (\ref{e6}) again for the paracathetal/paraxial approximation.

For reflection in a \textit{plane} mirror, we put ${r\!\to\!\infty}$ in the exact equation~(\ref{e9}), which then reduces to ${\sigma'\text{=\,\textminus}\sigma}${\tiny\,} for all~$\phi$, confirming that the image is stigmatic, on the cathetus, and as far behind the mirror as the object-point is in front. Later we shall find other uses for the exact equations (\ref{e7}) and~(\ref{e9}).

\subsection{Auxiliary axis}

From an object-point \textit{off} the axis of a coaxial system, a cathetus dropped to a facing spherical surface is not generally an axis of the whole system. But it is still an axis of that surface\textemdash wherefore it may be called an \textit{auxiliary axis}, while the axis of the system may be called the \textit{principal axis}\textemdash and a ray incident along that cathetus still offers the convenience of being undeviated by that surface. This convenience is exploited by Jenkins \&~White (\hyperlink{jenkins-white-76}{1976}) to find the image formed by refraction into a denser medium at a convex surface (p.\,51, Fig.\,3F) or a concave surface (p.\,52, Fig.\,3G), or by reflection at a concave surface (pp.\,100--101, Fig.\,6E) or a convex surface (p.\,101 \&~Fig.\,6F).\footnote{\,\small The last two examples are also given by Hecht (\hyperlink{hecht-17}{2017}, p.\,197, Fig.\,5.63), except that he does not use the term \textit{auxiliary~axis}, but explains the concept using ``Ray-1'' in his Fig.\,5.62 (p.\,196).} In each case, one ray is chosen to pass through the center of curvature\textemdash that~is, along the cathetus\textemdash and there are two candidates for a second ray, either of which (within the accuracy of the method) cuts the cathetus at the image-point.\footnote{\,\small{}The ``two candidates''\!, one incident parallel to the principal axis and the other refracted parallel to that axis, would be enough by themselves, especially as the authors (Jenkins \&~White, \hyperlink{jenkins-white-76}{1976}) are describing what they call the \textit{parallel-ray method}; but, idiosyncratically, they mention the undeviated ray before the second parallel ray (p.\,51, and again on p.\,101).}

For refraction at a single surface, as the same authors show (p.\,52 \&~Fig.\,3H), we can even use an auxiliary axis to locate the image of an object-point on the principal axis. First we construct the auxiliary axis parallel to the oblique incident ray from the object-point. This axis crosses the focal surface (which must be determined separately) at a point on the refracted oblique ray, fixing the direction of that ray, which then meets the principal axis at the desired image-point. In~effect, the cathetus rule is used twice\textemdash first to find the image of a hypothetical object-point at infinity, fixing the direction of a refracted ray from the actual object-point, and second to find the image of that point on the cathetus from that point.\footnote{\,\small{}In the corresponding case for a concave mirror (\hyperlink{jenkins-white-76}{Jenkins \&~White, 1976}, pp.\,101--2, Fig.\,6G), where the authors say ``If in place of ray~4 another ray were drawn through \textit{C} and parallel to ray~3,'' they are referring to an auxiliary axis, but they do not actually draw it.} The extension of the method to multiple surfaces is obvious.

The same authors, in a diagram already mentioned (p.\,48), show seven rays diverging from an object-point and refracted by a spherical surface to a real image-point, with one of the rays passing through the center of curvature but not otherwise labeled. In~the corresponding diagram for reflection (p.\,100, Fig.\,6C), the ray through the center of curvature is labeled the auxiliary axis, and all the other rays are shown as cutting this ray at the image-point. In~each case, the image as drawn (\textit{assumed} to be stigmatic) is located in accordance with the cathetus rule.

\subsection{Undeviated ray}

Wherever the cathetus rule holds\textemdash that~is, wherever the image is stigmatic and the cathetus well defined\textemdash the necessary and sufficient property of the cathetus is that \textit{a ray incident along the cathetus is undeviated}. Thus, if the image of an object-point is approximately stigmatic within a working aperture that admits an approximately undeviated ray, then, subject to those approximations, the image lies at the intersection of the undeviated ray and any other emergent ray (produced if necessary) from the same object-point. In~short, the approximately undeviated ray plays the role of the cathetus.

\begin{figure}[t]\centering
\includegraphics[width=3.74in]{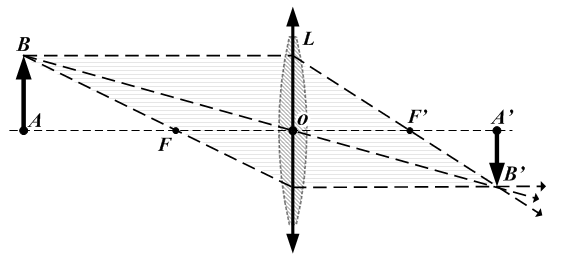}
\caption{Location of the image~$B'$ of an object-point~$B$ due to a thin lens. The (approximately) undeviated ray~$\mathit{BOB'}$ plays the role of the ancient cathetus: the image may be taken to be at the intersection of this ray and any other refracted ray originating at~$B$. (\href{https://commons.wikimedia.org/wiki/File:ThinLens.png}{Diagram by `Tamasflex'} at \textit{Wikimedia Commons}.)}
\end{figure}

A ray through the center of a \textit{thin} lens\textemdash that~is, a lens whose thickness is negligible compared with the object and image distances\textemdash may be considered undeviated even if it is oblique to the principal axis. This ray plays the same role in Newton's ``\textit{Cas}.\,4'' that the cathetus plays in his ``\textit{Cas}.\,2''\!.\footnote{\,\small\hyperlink{newton-2010}{Newton (2010)}, pp.\,11--13. More precisely, the nominated center is midway between the front and back focal points.} It~plays the same role in Fig.\,4C of Jenkins \&~White (\hyperlink{jenkins-white-76}{1976}, p.\,63) that the ray through the center of curvature plays in their Fig.\,3D (p.\,48), and (under the name ``chief ray'') the same role in their Figs.~4B, 4D, \&~4E (pp.\,62,\,63,\,64) that the cathetus respectively plays, anonymously in their Fig.\,3C (p.\,47) and as the ``auxiliary axis'' in their Figs.\,3F \&~3H (pp.\,51,\,53). More constructions reminiscent of the cathetus rule, with the ray through the center of the lens in the role of the cathetus, can be found in their Figs.~4F, 4G, 4H (for each lens), 4I (ditto), and 7B, and in (e.g.) Figs.~5.23, 5.24, and 5.29 of Hecht (\hyperlink{hecht-17}{2017}, pp.\,172,\,176).

For an object-point on the principal axis of the lens, the ray along that axis is \textit{exactly} undeviated and serves as the cathetus for the entire lens, so that the cathetus rule applies to the entire lens if the image is stigmatic. Examples of this sort (again not mentioning the cathetus rule) can be discerned in Fig.\,4A of Jenkins \&~White, and in Fig.\,5.15 of Hecht (\hyperlink{hecht-17}{2017},~p.\,168).

\section[Off-axis astigmatism\vspace{-1.2ex}]{Off-axis astigmatism}

The foregoing examples from Jenkins \&~White (\hyperlink{jenkins-white-76}{1976}) and Hecht (\hyperlink{hecht-17}{2017}) use \href{https://en.wikipedia.org/wiki/Gaussian optics}{Gaussian approximations}. They can model \href{https://en.wikipedia.org/wiki/Chromatic aberration}{chromatic aberration} if we allow for variation of refractive indices with wavelength. But if they are to model 3rd-order monochromatic aberrations in the \href{https://en.wikipedia.org/wiki/Meridional ray}{meridional} plane (\href{https://en.wikipedia.org/wiki/Spherical aberration}{spherical aberration}, tangential \href{https://en.wikipedia.org/wiki/Coma (optics)}{coma}, curvature of the tangential focal surface, and \href{https://en.wikipedia.org/wiki/Distortion (optics)}{distortion}), they must be modified\textemdash perhaps by resorting to trigonometric ray-tracing in the meridional plane,\footnote{\,\small See, e.g., \hyperlink{born-wolf-02}{Born \&~Wolf, 2002}, pp.\,204--7.} in which case we still have the problem of assessing aberrations that involve rays outside that plane. For sagittal coma we can use the well-known proportionality (to leading order) between sagittal and tangential coma.\footnote{\,\small\hyperlink{jenkins-white-76}{Jenkins \&~White, 1976}, p.\,164.} For \href{https://en.wikipedia.org/wiki/Astigmatism (optical systems)}{astigmatism}, however, we need a sample of rays outside the meridional plane. With spherical surfaces, the easiest way to take such a sample is to exploit the exactness of the cathetus rule for the sagittal image formed by a surface axially symmetrical about the cathetus. And this is where we reap the reward for delaying approximations in the above derivation of the ``Gaussian formula''\!.

In our \hyperref[f-sagit]{Figure~{\tiny\!}6}, suppose that the line $\mathit{OC}$ is \textit{not} the principal axis, but only an auxiliary axis. Let~$O$ be an off-axis object-point or an intermediate image thereof; and from~$O$, let $\mathit{OPI}$ be the path of the \textbf{chief~ray}\textemdash that~is, the ray through the center of the main aperture (wherever the main aperture stop happens to be). Then the sagittal image formed by the surface~$\mathit{VP}$ is~$I$, whose position is given by equation~(\ref{e7}) for a refractive surface, or~(\ref{e9}) for a reflective surface. Equivalent results are given by Jenkins \&~White, citing the derivation by Monk,\footnote{\,\small\hyperlink{jenkins-white-76}{Jenkins \&~White, 1976}, p.\,169, Eqs.\,(9p), 2nd eq.~(for refraction) and p.\,111, 2nd eq.~(for reflection), citing \hyperlink{monk-63}{Monk, 1963}, pp.\,424--6.} who begins by saying that ``if coma is absent, all the rays which have the same inclination\ldots\,as~$\mathit{OP}$ with~$\mathit{OC}$ will intersect the line~$\mathit{OC}$\ldots\,in a point'' which we call~$I$. The condition that ``coma is absent'' is redundant because the conclusion follows from the axial symmetry about~$\mathit{OC}$ (which Monk ignores, calling $\mathit{PC}$ the axis). No such condition is assumed in the earlier derivation by \href{https://en.wikipedia.org/wiki/Alexander Eugen Conrady}{Conrady}, first published in 1929,\footnote{\,\small\hyperlink{conrady-92}{Conrady (1992)}, pp.\,409--10.} which duly invokes the auxiliary axis, and which, in~spite of its different sign convention, is the main source for our derivation of (\ref{e7}) above. Conrady's equation (d) corresponds to our~(\ref{e7}), and agrees with the result that \href{https://en.wikipedia.org/wiki/Principles of Optics}{Born \&~Wolf} obtain by a longer process, involving a ``thin pencil'' of rays and a Hamiltonian characteristic function.\footnote{\,\small\hyperlink{born-wolf-02}{Born \&~Wolf, 2002}, p.\,186, Eq.\,(22).} None of these sources uses the word \textit{cathetus} or refers to the cathetus rule.

Corresponding expressions for the distance of the \textit{tangential} image along~$\mathit{PI}$ are given by the same authors and\textemdash most remarkably\textemdash by Barrow.\footnote{\,\small On Barrow, and Newton's deference to him in this matter, see \hyperlink{shapiro-1990}{Shapiro, 1990}, pp.\,135--6,\,147--8, and \hyperlink{newton-anon-1728}{Newton/anon., 1728}, p.\,107.} In~principle, we locate the tangential image by moving~$P$ along the arc~$\mathit{VP}$ (by an infinitesimal distance if we want an analytical result, or a finite distance if we are tracing rays numerically) and finding the intersection of the new~$\mathit{PI}$ with the old. The distance between the tangential and sagittal images along the old~$\mathit{PI}$ is a measure of the astigmatism.

By the axial symmetry, as we scan the aperture by rotating the arc~$\mathit{VP}$ about the axis~$\mathit{OC}$, the tangential image likewise rotates about that axis, tracing a circular arc; and as we scan the aperture by moving~$P$ away from~$V$, the sagittal image can only move along that axis. So the tangential and sagittal focal lines are perpendicular to each other; but the sagittal focal line is \textit{not} generally perpendicular to the chief ray~$\mathit{PI}$ (although the tangential focal line~is). Thus, as Born \&~Wolf note, it is not generally true that the focal lines are perpendicular to the chief ray ``as is often incorrectly asserted in the literature''\!.\footnote{\,\small\hyperlink{born-wolf-02}{Born \&~Wolf, 2002}, p.\,182. Earlier on the same page, Born \&~Wolf themselves may seem to have asserted what they now deny. But the exculpatory words are ``To the first order''; for a \textit{thin} pencil, if the distance between the focal lines measured along the central ray is first-order, then the obliquity of either focal line to the central ray is \textit{second}-order.} Indeed I~have noticed that the offenders include Jenkins \&~White (\hyperlink{jenkins-white-76}{1976}), who claim that the sagittal focal line, which they call~\textit{S}, is perpendicular to what they call the \textit{sagittal plane} (p.\,169), which contains the chief ray and is perpendicular to the \textit{tangential plane} (meridional plane; see their Fig.\,9P). They go on to say that on the sagittal focal surface, the images are ``parallel to the spokes'' (p.\,169), whereas in fact the sagittal focal line for a point on a spoke need only be in the plane of the spoke and the axis. Their Fig.\,6N (p.\,112) is similarly misleading; the sagittal focal line \textit{S} should be along the auxiliary axis\textemdash that~is, parallel to the incoming rays (the object-point being at infinity).

In our \hyperref[f-sagit]{Figure~{\tiny\!}6}, the \textit{sagittal plane} after refraction is the plane perpendicular to the plane of the diagram and containing the ray~$\mathit{PI}$. If we leave the sagittal plane fixed and rotate the point of refraction about the axis (cathetus)~$\mathit{OC}$, the circle traced on the refracting surface is not identical to the intersection of that surface with the sagittal plane, but is tangential to that intersection, and the tangency is enough for calculating the astigmatism to leading order.\footnote{\,\small Compare the corresponding remarks by Conrady (\hyperlink{conrady-92}{1992}, top of p.\,410).} Thus Born \&~Wolf get the same sagittal equation as Conrady in spite of their radically different method. In~a coaxial system, as $P$ traces a circle with axis~$\mathit{OC}$, the path traced by the intersection of the refracted ray~$\mathit{PI}$ with the \textit{next} surface is not generally a circle with its axis on the cathetus from $I$ to that surface, but again is tangential to such a circle. Hence equation (\ref{e7}) or~(\ref{e9}) can be used with successive surfaces to find the successive positions of the sagittal image on the chief ray, and assess the final astigmatism, to leading order.

\section[Conclusion: Unreasonable in what sense?\vspace{-1.2ex}]{Conclusion: Unreasonable in what sense?}

It has been shown that there are conditions under which the cathetus rule is true or nearly so. Let it be conceded that under these conditions the rule must be, in some sense, effective, and that this effectiveness, as far as it goes, is by definition reasonable. One might object that these conditions\textemdash that the image is stigmatic or nearly so, and the cathetus unambiguous\textemdash seem narrow, and that the effectiveness of the rule, by comparison, seems unreasonably wide. In~response, one could point out that surfaces forming stigmatic or nearly stigmatic images are useful and therefore likely to be encountered in practice, and likely to encourage propagation of any principle found applicable to them. Moreover, the shapes nominated by Newton as producing nearly stigmatic images\textemdash plane or spherical, or, let us add, nearly so\textemdash may exist for reasons unrelated to their imaging properties: I~may see my face reflected in a teapot, though the teapot is not an optical device. For these reasons, examples of the effectiveness of the rule might reasonably be prevalent, or at least prominent.

When we delve into the history of that ``antient and common'' rule, however, any semblance of reasonableness evaporates.

The cathetus rule was unanimously upheld for nearly 19 centuries although there was not a single non-tautological case in which the rule had been validly demonstrated. Even the tautological case\textemdash that in which the line of sight is along the cathetus\textemdash was botched from the beginning (recall Euclid's ``postulates''), and eventually put on a secure footing after 13 centuries when Alhacen posed the examples of the eye lining up a sharp tip with its reflection, and the eye looking at its own reflection.

But, after Kepler's attack in 1604 sent the rule into decline, only one more century passed before the rule was rehabilitated, without acknowledgment, by Newton's widely applicable ``axiom'' of approximate stigmatism, whereby the cathetus\textemdash disguised as the axis or the auxiliary axis or (generalized) as the undeviated ray\textemdash made itself extremely useful in ``Gaussian'' optics. Meanwhile the \textit{exact} application of the rule to the sagittal image, for axial symmetry about the cathetus, languished in Newton's posthumous lecture notes, but reappeared in the 20th century\textemdash unnamed and unsourced\textemdash for the evaluation of 3rd-order astigmatism in coaxial systems with spherical surfaces, yielding the same formula as Hamiltonian theory, with less labor and less conceptual difficulty.

For nearly nineteen centuries, until Benedetti (1585), the cathetus rule was a non-sequitur: the effectiveness of the rule, in so far as it was correctly described, was unreasonably unexplained. For the three centuries since Newton, it has been unreasonably unrecognized.

\section[Acknowledgments\vspace{-1.2ex}]{Acknowledgments}

If my analysis of Benedetti (\hyperlink{benedetti-1585}{1585}) adds any value to Goulding's (\hyperlink{goulding-18}{2018}), much of the credit is due to \href{https://en.wikipedia.org/wiki/Google Translate}{Google Translate} and \href{https://en.wikipedia.org/wiki/ChatGPT}{ChatGPT}~3.5; the latter (with a few ``custom instructions'') expedited the correction of \href{https://en.wikipedia.org/wiki/Optical character recognition}{OCR} errors in the plain text from \href{https://en.wikipedia.org/wiki/Google Books}{Google Books}, and then gave a second opinion on translation.

The \href{https://en.wikiversity.org/wiki/WikiJournal_Preprints/The_unreasonable_effectiveness_of_the_cathetus_rule_in_ancient_and_modern_optics}{\textit{WikiJournal} version}{\tiny\,} of this paper has separate ``Notes'' and ``Citations''\!, which this \textit{arXiv} version combines as sequentially numbered footnotes.

\section[Further reading\vspace{-1.2ex}]{Further reading}

Of all the authors cited above, Goulding (\hyperlink{goulding-18}{2018}), although his account ends early in the 17th century, has by far the most information on the cathetus rule (including much detail on the modifications by Brengger and Stevin), and he alone reports Benedetti's priority in disproving and salvaging the~rule.

For a concise general history of optics over the life of the traditional cathetus rule, see A.\,Mark Smith, ``Optics to the time of Kepler''\!, \textit{Encyclopedia of the History of Science} (Nov.\,2022; rev.~Jul.\,2023), \href{https://doi.org/10.34758/v9kd-ad56}{doi.org/10.34758/v9kd-ad56}. For a more expansive version, see \hyperlink{smith-2017}{Smith, 2017}.

\section{Bibliography}\small\raggedright

\parskip 0.5ex
\leftskip 15pt
\parindent -15pt

\hspace{\parindent}\hypertarget{bacon-burke-1928}{R.~Bacon, tr.~R.B.\,Burke, 1928, \textit{The Opus Majus of Roger Bacon} (2~vols.), University of Pennsylvania Press, vol.\,2.}

\hypertarget{bacon-combach-1614}{R.~Bacon (ed.~J.\,Combach), 1614, \textit{Perspectiva}, Frankfurt: Wolfgang Richter for Anton Humm; \href{https://books.google.com/books?id=Cn6k7IC-yaMC}{google.com/books?id=Cn6k7IC-yaMC}.}

\hypertarget{barrow-1669}{I.~Barrow, 1669, \textit{Lectiones XVIII, Cantabrigi\ae\ in scholis publicis habit\ae; in quibus opticorum ph\ae{}nomen$\omega$n genuin\ae\ rationes investigantur, ac exponuntur}, London: William Godbid; \href{https://books.google.com/books?id=WpB_5y0XcN4C}{google.com/books?id=WpB\_5y0XcN4C}.}

\hypertarget{barrow-fay-87}{I.~Barrow, tr.~H.C.\,Fay, 1987, \textit{Isaac Barrow's Optical Lectures} (ed.~A.G.\,Bennett \& D.F.~Edgar), London: Worshipful Company of Spectacle Makers.}

\hypertarget{benedetti-1585}{G.B.~Benedetti, 1585, \textit{Diversarum Speculationum Mathematicarum, et Physicarum, Liber}, Turin: Heirs of Nicol\`{o} Bevilacqua; \href{https://books.google.com/books?id=Ec6bHphLvzMC}{google.com/books?id=Ec6bHphLvzMC} / \href{https://books.google.com/books?id=lhOWpKH6I_MC}{google.com/books?id=lhOWpKH6I\_MC}.}

\hypertarget{berkeley-1901}{G.~Berkeley (1901), ``An essay towards a new theory of vision''\!, 1709--32, in~A.C.\,Fraser~(ed.), \textit{The~Works of George Berkeley D.D.} (4~vols.), Oxford, 1901, vol.\,1, \href{https://archive.org/details/worksofberkeley01berkuoft}{archive.org/details/worksofberkeley01berkuoft}, pp.\,121--210.}

\hypertarget{born-wolf-02}{M.~Born and E.\,Wolf, 2002, \textit{Principles of Optics}, 7th~Ed., Cambridge, 1999 (reprinted with corrections, 2002).}

\hypertarget{cardona-gutierrez-20}{C.A.\,Cardona and J.\,Guti\'{e}rrez, 2020, ``On Berkeley's solution to the Barrovian case''\!, \textit{Principia: An~International Journal of Epistemology}, vol.\,24, no.\,2, pp.\,363--89; \href{https://doi.org/10.5007/1808-1711.2020v24n2p363}{doi.org/10.5007/1808-1711.2020v24n2p363} (\href{https://periodicos.ufsc.br/index.php/principia/article/view/64997/44662}{PDF}, open access).}

\hypertarget{cavalieri-1647}{B.~Cavalieri, 1647, \textit{Exercitationes Geometricae Sex}, Bologna: Giacomo Monti; \href{https://archive.org/details/bub_gb_OXe4dPXGSDMC}{archive.org/details/bub\_gb\_OXe4dPXGSDMC}.}

\hypertarget{conrady-92}{A.E.~Conrady (1992), \textit{Applied Optics and Optical Design}, Part~1 (first published London: Oxford University Press, 1929), New\,York: Dover, 1957, 1985, 1992.}

\hypertarget{darrigol-12}{O.~Darrigol, 2012, \textit{A History of Optics: From Greek Antiquity to the Nineteenth Century}, Oxford.}

\hypertarget{dijksterhuis-55}{E.J.~Dijksterhuis (ed.), 1955, \textit{The Principal Works of Simon Stevin}, vol.\,1, Amsterdam: Swets \& Zeitlinger.}

\hypertarget{dijksterhuis-99}{F.J.~Dijksterhuis, 1999, \textit{Lenses and Waves: Christiaan Huygens and the Mathematical Science of Optics in the Seventeenth Century} (doctoral thesis), University of Twente; \href{http://doc.utwente.nl/33764/}{doc.utwente.nl/33764}. (See~also the book with the~same author and title,\, Dordrecht: Kluwer Academic Publishers, 2004.)}

\hypertarget{dijksterhuis-04}{F.J.~Dijksterhuis, 2004, ``Once Snell breaks down: From geometrical to physical optics in the seventeenth century''\!, \textit{Annals of Science}, vol.\,61, no.\,2 (Apr.\,2004), pp.\,165--85.}

\hypertarget{dupre-10}{S.~Dupr\'{e}, 2010, ``William Bourne's invention.~Projecting a telescope and optical speculation in Elizabethan England''\!, in~\hyperlink{vanHelden-et-al-10}{Van Helden et~al., 2010}, pp.\,129--45.}

\hypertarget{euclid-dasypodius-1557}{Euclid (attrib.) and C.\,Dasypodius (tr.), 1557, ${\mathit{E\upsilon\kappa\lambda\epsilon\iota\delta{}o\upsilon}}$ ${\mathit{K\alpha\tau\acute{o}\pi\tau\varrho\iota\kappa\alpha}}$ / \textit{Euclidis Catoptrica} (Greek~and Latin, in~facing columns, with no page numbers), Strasbourg (``Argentorati''): Rihel; \href{https://books.google.com/books?id=r4Y8AAAAcAAJ}{google.com/books?id=r4Y8AAAAcAAJ}.}

\hypertarget{euclid-heiberg-1895}{Euclid (attrib.) and J.L.\,Heiberg (ed.), 1895, \textit{Catoptrica} (Greek and Latin, on facing pages), in~J.L.\,Heiberg \& H.\,Menge (eds.), \textit{Euclidis Opera Omnia} (9~vols.), Leipzig: Teubner, 1883--99, vol.\,7, \href{https://books.google.com/books?id=0-0XAAAAMAAJ}{google.com/books?id=0-0XAAAAMAAJ}, pp.\,285--343.}

\hypertarget{euclid-pena-1557}{Euclid (attrib.) and J.\,Pena (tr.), 1557, ${\mathit{E\upsilon\kappa\lambda\epsilon\acute{\iota}\delta{}o\upsilon}}$ ${\mathit{O\pi\tau\iota\kappa\grave{\alpha}}}$ ${\mathit{\kappa\alpha\grave{\iota}}}$ ${\mathit{K\alpha\tau{}o\pi\tau\varrho\iota\kappa\acute{\alpha}}}$ / \textit{Euclidis Optica \& Catoptrica} (in~Greek and Latin), Paris: Andreas Wechelus; \href{https://books.google.com/books?id=StJOmA9SOTsC}{google.com/books?id=StJOmA9SOTsC}.}

\hypertarget{goulding-18}{R.~Goulding, 2018, ``Binocular vision and image location before Kepler''\!, \textit{Archive for History of Exact Sciences}, vol.\,72, no.\,5 (Sep.\,2018), pp.\,497--546; \href{https://www.jstor.org/stable/45211958}{jstor.org/stable/45211958}.}

\hypertarget{goulding-22}{R.~Goulding, 2022, ``The harvest of optics: Descartes, Mydorge, and their paths to a theory of refraction''\!, \textit{Annals of Science}, vol.\,79, no.\,2, pp.\,164--214 (March~2022); \href{https://doi.org/10.1080/00033790.2022.2026479}{doi.org/10.1080/00033790.2022.2026479}.}

\hypertarget{gregory-1663}{J.~Gregory, 1663, \textit{Optica Promota}, London: J.\,Hayes; \href{https://books.google.com/books?id=W2U_AAAAcAAJ}{google.com/books?id=W2U\_AAAAcAAJ}.}

\hypertarget{gregory-bruce-06}{J.~Gregory, and I.\,Bruce (tr.\,\&\,ed.), 2006, \textit{James Gregory's OPTICA PROMOTA} (in~English and Latin), \href{http://17centurymaths.com/contents/contentsGregory.htm}{17centurymaths.com/contents/contentsGregory.htm}.}

\hypertarget{halliwell-1839}{J.O.~Halliwell (ed.), 1839, \textit{Rara Mathematica; or, A collection of treatises on the mathematics and subjects connected with them, from ancient inedited manuscripts}, London: John William Parker, \href{https://archive.org/details/raramathematicao00hallrich}{archive.org/details/raramathematicao00hallrich}.}

\hypertarget{hecht-17}{E.~Hecht, 2017, \textit{Optics}, 5th (Global) Ed., Pearson Education.}

\hypertarget{jenkins-white-76}{F.A.\,Jenkins and H.E.\,White, 1976, \textit{Fundamentals of Optics}, 4th~Ed., New\,York: McGraw-Hill.}

\hypertarget{kepler-1604}{J.~Kepler, 1604, \textit{Ad Vitellionem Paralipomena, quibus Astronomiae Pars Optica traditur}, Frankfurt: Claude de~Marne \& the heirs of Johann Aubry; \href{https://books.google.com/books?id=vLUgK23vc90C}{google.com/books?id=vLUgK23vc90C}.}

\hypertarget{kepler-1859}{J.~Kepler (1859), \textit{Dioptrice} (first published Augsburg: Davidus Francus, 1611), in~Kepler, \textit{Opera~Omnia} (ed.~C.\,Frisch), vol.\,2, Frankfurt \& Erlangen: Heyder \& Zimmer, 1859, \href{https://books.google.com/books?id=KG6qKr4Ln90C}{google.com/books?id=KG6qKr4Ln90C}, pp.\,515--74.}

\hypertarget{kepler-donahue-00}{J.~Kepler, tr.~W.H.\,Donahue, 2000, \textit{Optics: Paralipomena to Witelo \& Optical Part of Astronomy}, Santa~Fe: Green Lion Press (reprinted 2021).}

\hypertarget{lindberg-71}{D.C.~Lindberg, 1971, ``Lines of influence in thirteenth-century optics: Bacon, Witelo, and Pecham''\!, \textit{Speculum}, vol.\,46, no.\,1 (Jan.\,1971), pp.\,66--83; \href{https://www.jstor.org/stable/2855089}{jstor.org/stable/2855089}.}

\hypertarget{lindberg-81}{D.C.~Lindberg, 1981, \textit{Theories of vision from al-Kindi to Kepler}, University of Chicago Press, 1976 (paperback~ed., 1981).}

\hypertarget{lohne-59}{J.~Lohne, 1959, ``Thomas Harriott (1560--1621): The Tycho Brahe of optics''\!, \textit{Centaurus}, vol.\,6, no.\,2 (June~1959), pp.\,113--121.}

\hypertarget{malet-1990g}{A.~Malet, 1990g, ``Gregorie, Descartes, Kepler, and the law of refraction''\!, \textit{Archives Internationales d'Histoire des Sciences}, vol.\,40 (1990), no.\,125, pp.\,278--304.}

\hypertarget{malet-1990k}{A.~Malet, 1990k, ``Keplerian illusions: Geometrical pictures \textit{vs} optical images in Kepler's visual theory''\!, \textit{Studies~in History and Philosophy of Science}, vol.\,21, no.\,1 (March~1990), pp.\,1--40; \href{https://doi.org/10.1016/0039-3681(90)90013-X}{doi.org/10.1016/0039-3681(90)90013-X}.}

\hypertarget{malet-2003}{A.~Malet, 2003, ``Kepler and the telescope''\!, \textit{Annals of Science}, vol.\,60, no.\,2, pp.\,107--36; \href{https://doi.org/10.1080/0003379031000080961}{doi.org/10.1080/0003379031000080961}.}

\hypertarget{malet-2010}{A.~Malet, 2010, ``Kepler's legacy: telescopes and geometrical optics, 1611--1669''\!, in~\hyperlink{vanHelden-et-al-10}{Van Helden et~al., 2010}, pp.\,281--300; \href{https://upf.academia.edu/AntoniMalet}{upf.academia.edu/AntoniMalet}.}

\hypertarget{monk-63}{G.S.~Monk, 1963, \textit{Light: Principles and Experiments}, 2nd~Ed., New\,York: Dover.}

\hypertarget{newton-2010}{I.~Newton (2010), \textit{Opticks: or, a Treatise of the Reflections, Refractions, Inflections, and Colours of Light}, 4th~Ed.~(first published London: William Innys, 1730), Mineola, NY: Dover, 1952, 1979, 2012; Project Gutenberg, 2010, \href{http://www.gutenberg.org/ebooks/33504}{gutenberg.org/ebooks/33504}. (Cited page numbers match the Gutenberg {\footnotesize HTML} editions and the Dover editions.)}

\hypertarget{newton-anon-1728}{I.~Newton, tr.~anon., 1728, \textit{Optical Lectures Read in the Publick Schools of the University of Cambridge, Anno Domini, 1669} [sic], London: Francis Fayram; \href{https://archive.org/details/bim_eighteenth-century_lectiones-opticae-engli_newton-sir-isaac_1728}{archive.org/details/bim\_eighteenth-century\_lectiones-opticae-engli\_newton-sir-isaac\_1728}.}

\hypertarget{pecham-gaurico-1504}{J.~Pecham (ed.~L.\,Gaurico), 1504, \textit{Perspectiva Communis}, Venice: Giovan Battista Sessa; \href{https://archive.org/details/joarchiepiscopic00peck}{archive.org/details/joarchiepiscopic00peck}. (The publisher's mark appears below the frontispiece.)}

\hypertarget{pecham-hartmann-1542}{J.~Pecham (ed.~G.\,Hartmann), 1542, \textit{Perspectiva Communis}, Nuremberg: Johannes Petreius; \href{https://archive.org/details/perspectivacommv00peck}{archive.org/details/perspectivacommv00peck}.}

\hypertarget{ptolemy-govi-1885}{C.~Ptolemy (ed.~G.\,Govi), 1885, \textit{L'Ottica di Claudio Tolomeo} (in~Latin; introduction in Italian), Turin: Paravia; \href{https://archive.org/details/lotticadiclaudi00eugegoog}{archive.org/details/lotticadiclaudi00eugegoog}.}

\hypertarget{risner-1572}{F.~Risner (ed.), 1572, \textit{Opticae Thesaurus.~Alhazeni Arabis libri septem, nunc prim\`{u}m editi\ldots Vitellonis Thuringolopoli libri~X} (one vol.; two parts, separately paginated), Basel: per Episcopios; \href{https://books.google.com/books?id=V27nL0HJd78C}{google.com/books?id=V27nL0HJd78C}.}

\hypertarget{rohault-clarke-1735}{J.~Rohault (ed.~S.\,Clarke, tr.~J.\,Clarke), 1735, \textit{Rohault's System of Natural Philosophy}, 3rd~Ed., London: Knapton, vol.\,1; \href{https://archive.org/details/b30535578_0001}{archive.org/details/b30535578\_0001}.}

\hypertarget{sabra-67}{A.I.~Sabra, 1967, ``The authorship of the \textit{Liber de crepusculis}, an eleventh-century work on atmospheric refraction''\!, \textit{Isis}, vol.\,58, no.\,1 (Spring 1967), pp.\,77--85; \href{https://www.jstor.org/stable/228388}{jstor.org/stable/228388}.}

\hypertarget{schuster-00}{J.A.~Schuster, 2000, ``Descartes \textit{opticien}: The construction of the law of refraction and the manufacture of its physical rationales, 1618--29''\!, in~S.\,Gaukroger, J.A.\,Schuster, \& J.\,Sutton (eds.), \textit{Descartes' Natural Philosophy}, London: Routledge, pp.\,258--312.}

\hypertarget{shapiro-1990}{A.E.~Shapiro, 1990, ``The \textit{Optical Lectures} and the foundations of the theory of optical imagery''\!, in~M.\,Feingold~(ed.), \textit{Before Newton: The Life and Times of Isaac Barrow}, Cambridge, pp.\,105--78.}

\hypertarget{shapiro-2008}{A.E.~Shapiro, 2008, ``Images: Real and Virtual, Projected and Perceived, from Kepler to Dechales''\!, \textit{Early Science and Medicine}, vol.\,13, no.\,3, pp.\,270--312; \href{https://www.jstor.org/stable/20617731}{jstor.org/stable/20617731}.}

\hypertarget{shirley-51}{J.W.~Shirley, 1951, ``An early experimental determination of Snell's law''\!, \textit{American Journal of Physics}, vol.\,19, no.\,9 (Dec.\,1951), pp.\,507--8; \href{https://doi.org/10.1119/1.1933068}{doi.org/10.1119/1.1933068}.}

\hypertarget{smith-1996}{A.M.~Smith (tr.), 1996, ``Ptolemy's theory of visual perception: An~English translation of the \textit{Optics} with introduction and commentary''\!, \textit{Transactions of the American Philosophical Society}, vol.\,86, no.\,2; \href{https://www.jstor.org/stable/3231951}{jstor.org/stable/3231951}.}

\hypertarget{smith-2001}{A.M.~Smith (tr.\,\&\,ed.), 2001, ``Alhacen's theory of visual perception: A~critical edition, with English translation and commentary, of the first three books of Alhacen's \textit{De\,Aspectibus}, the medieval Latin version of Ibn\,al-Haytham's \textit{Kit\={a}b al-Man\={a}zir}''\!, in~\textit{Transactions of the American Philosophical Society}, vol.\,91, no.\,4, \href{https://www.jstor.org/stable/3657358}{jstor.org/stable/3657358} (vol.\,1: Introduction and Latin text), and no.\,5, \href{https://www.jstor.org/stable/3657357}{jstor.org/stable/3657357} (vol.\,2: English translation).}

\hypertarget{smith-2006}{A.M.~Smith (tr.\,\&\,ed.), 2006, ``Alhacen on the principles of reflection: A~critical edition, with English translation and commentary, of Books 4 and 5 of Alhacen's \textit{De\,Aspectibus}, the medieval Latin version of Ibn\,al-Haytham's \textit{Kit\={a}b al-Man\={a}zir}''\!, in~\textit{Transactions of the American Philosophical Society}, vol.\,96, no.\,2, \href{https://www.jstor.org/stable/20020399}{jstor.org/stable/20020399} (vol.\,1: Introduction and Latin text), and no.\,3, \href{https://www.jstor.org/stable/20020403}{jstor.org/stable/20020403} (vol.\,2: English translation).}

\hypertarget{smith-2008}{A.M.~Smith (tr.\,\&\,ed.), 2008, ``Alhacen on image-formation and distortion in mirrors: A~critical edition, with English translation and commentary, of Book~6 of Alhacen's \textit{De\,Aspectibus}, the medieval Latin version of Ibn\,al-Haytham's \textit{Kit\={a}b al-Man\={a}zir}'' (vol.\,2: English translation), \textit{Transactions of the American Philosophical Society}, vol.\,98, no.\,1, sec.\,2; \href{https://www.jstor.org/stable/27757399}{jstor.org/stable/27757399}.}

\hypertarget{smith-2010}{A.M.~Smith (tr.\,\&\,ed.), 2010, ``Alhacen on Refraction: A~critical edition, with English translation and commentary, of Book~7 of Alhacen's \textit{De\,Aspectibus}, the medieval Latin version of Ibn\,al-Haytham's \textit{Kit\={a}b~al-Man\={a}zir}'' (vol.\,2: English translation), \textit{Transactions of the American Philosophical Society}, vol.\,100, no.\,3, sec.\,2; \href{https://www.jstor.org/stable/20787651}{jstor.org/stable/20787651}.}

\hypertarget{smith-2017}{A.M.~Smith, 2017, \textit{From Sight to Light: The Passage from Ancient to Modern Optics}, University of Chicago Press, 2015 (paperback ed., 2017).}

\hypertarget{tacquet-1669}{A.\,Tacquet, 1669, \textit{Catoptrica Tribus Libris Exposita}, in~\textit{Opera Mathematica}, Antwerp: Meursius, vol.\,2, \href{https://books.google.com/books?id=XHK2NgG3UfQC}{google.com/books?id=XHK2NgG3UfQC}, pp.\,213--264ff.}

\hypertarget{takahashi-92}{K.\,Takahashi, 1992, \textit{The Medieval Latin Traditions of Euclid's} Catoptrica: \textit{A~Critical Edition of}{\tiny\,} De speculis \textit{with an Introduction, English Translation and Commentary}, Kyushu University Press.}

\hypertarget{unguru-72}{S.~Unguru, 1972, ``Witelo and thirteenth-century mathematics: An assessment of his contributions''\!, \textit{Isis}, vol.\,63, no.\,4 (Dec.\,1972), pp.\,496--508; \href{https://www.jstor.org/stable/229773}{jstor.org/stable/229773}.}

\hypertarget{vanHelden-et-al-10}{A.\,Van~Helden, S.\,Dupr\'{e}, R.\,van~Gent, \& H.\,Zuidervaart (eds.), 2010, \textit{The Origins of the Telescope}, Amsterdam: {\footnotesize KNAW }Press; \href{https://dspace.library.uu.nl/handle/1874/224188}{dspace.library.uu.nl/handle/1874/224188} (open~access).}

\hypertarget{vollgraff-1936}{J.A.\,Vollgraff, 1936, ``Snellius' notes on the reflection and refraction of rays''\!, \textit{Osiris}, vol.\,1 (Jan.\,1936), pp.\,718--25; \href{https://www.jstor.org/stable/301634}{jstor.org/stable/301634}.}

\end{document}